\documentclass[a4paper,11pt]{article}

\usepackage{jheppub} 

\usepackage[T1]{fontenc} 

\usepackage{tikz-cd}
\usetikzlibrary{cd}

\newcommand{\br}{\color{red}}

\newcommand{\To}{\longrightarrow}

\newcommand \ket[1]{|\,{#1}\,\rangle}
\newcommand \bracket[2]{\langle\,{#1}\,|\,{#2}\,\rangle}

\title{\boldmath Classical conformal blocks, Coulomb gas integrals and Richardson--Gaudin models}

\author[a,1]{M.R.~Pi\c{a}tek,\note{Corresponding author.}}
\author[b]{R.G.~Nazmitdinov,}
\author[c]{A.~Puente}
\author[]{and A.R.~Pietrykowski}

\affiliation[a]{Institute of Physics, University of Szczecin,\\ Wielkopolska 15, 70-451 Szczecin, Poland}
\affiliation[b]{Bogoliubov Laboratory of Theoretical Physics, Joint Institute for Nuclear Research,\\ 141980 Dubna, Russia}
\affiliation[c]{Departament de Fisica, Universitat de les Illes Balears,\\ E-07122 Palma de Mallorca, Spain}

\emailAdd{marcin.piatek@usz.edu.pl}
\emailAdd{rashid@theor.jinr.ru}
\emailAdd{puente.toni@gmail.com}
\emailAdd{hearthie@gmail.com}

\abstract{
Virasoro conformal blocks are universal ingredients of correlation functions 
of two-dimensional conformal field theories (2d CFTs) with Virasoro symmetry. It 
is acknowledged that in the (classical) limit of large central charge of the 
Virasoro algebra and large external, and 
intermediate conformal weights with fixed ratios of these parameters 
Virasoro blocks exponentiate to functions known as 
Zamolodchikovs' classical blocks. The latter are special functions 
which have awesome mathematical and physical applications. 
Uniformization, monodromy problems, 
black holes physics, quantum gravity, entanglement, quantum chaos, 
holography, ${\cal N}=2$ gauge theory and quantum integrable systems (QIS) are 
just some of contexts, where classical Virasoro blocks are in use. 
In this paper, exploiting known connections between power 
series and integral representations of (quantum) 
Virasoro blocks, we propose new finite closed formulae for certain 
multi-point classical Virasoro blocks on the sphere. Indeed, combining 
classical limit of Virasoro blocks expansions with a saddle point 
asymptotics of Dotsenko--Fateev (DF) integrals one can relate 
classical Virasoro blocks with a critical value of the 
``Dotsenko--Fateev matrix model action''. The latter is the ``DF action'' 
evaluated on a solution of saddle point equations which take the form of 
Bethe equations for certain QIS (Gaudin spin models). 
A link with integrable models is our main motivation for 
this research line. For instance, analogous quantities as the ``DF on-shell 
action'' appear in 2d CFT realization of the so-called Richardson's solution of the reduced BCS 
model describing physics of ultra-small superconducting grains. 
Precisely, the Richardson solution and its particular limit characterizing the rational 
Gaudin model have known implementation in 2d CFT in 
terms of a free field representation of certain (perturbed and unperturbed) WZW blocks. 
The WZW analogue of the ``DF on-shell action'' appears here and 
plays a crucial role, it is a  generating function for eigenvalues of 
quantum integrals of motion. An exploration of 
relationships between power expansions and  Coulomb gas representations of 
the aforementioned WZW blocks could pave a way for new analytical 
tools useful in the study of models of the Richardson--Gaudin type.
}

\begin{document} 
\maketitle
\flushbottom

\section{Introduction}
Conformal field theory in two dimensions (2d CFT or CFT$_2$)
yields an effective description of critical phenomena 
in two-dimensional statistical systems \cite{DiFMS} 
and describes worldsheet dynamics of relativistic strings \cite{Pol}. 
Two-dimensional CFT has also found use in the condensed 
matter physics, in particular as a tool to study states of the fractional quantum Hall effect (FQHE).
For instance, the Laughlin wave
functions \cite{Laughlin} can be represented by certain conformal 
blocks calculated within a free field realization.\footnote{
For a review of applications of 2d CFT in theoretical investigations of the quantum Hall effect, see e.g.~\cite{HHSV}.}
Conformal blocks (see below) are model independent 2d CFT ``special functions'' defined entirely within a 
representation theoretic framework. 

During the last decade CFT$_2$ has been used in completely new contexts.
Before mentioning them, let us recall that in 2d CFT, like in any quantum field theory, basic objects 
being studied are correlation functions.
In 2d CFT correlation functions are defined on the Riemann surfaces $C_{g,n}$ with genus $g$ and $n$ punctures.
Due to conformal symmetry correlators of any 2d CFT model can be written as a sum (or an integral for 
theories with a continuous spectrum) which includes terms consisting of conformal blocks 
times the so-called structure constants.
The conformal blocks 
${\cal F}\!\left({\boldsymbol\Delta},\underline{\boldsymbol\Delta},c\,|\,\cdot\,\right)$
on $C_{g,n}$ 
depend on the central charge $c$ of the Virasoro algebra, the so-called external conformal weights 
$\boldsymbol\Delta:=\left\lbrace\Delta_i\right\rbrace_{i=1}^{n}$
the weights 
$\underline{\boldsymbol\Delta}:=\left\lbrace{\underline\Delta}_p\right\rbrace_{p=1}^{3g-3+n}$
in the intermediate 
channels, the vertex operators 
locations and modular parameters in case of surfaces with $g>0$.
Conformal blocks are fully determined by the underlying conformal symmetry. These functions have 
an interesting analytic structure, although it is not completely understood yet. 
In general, they can be expressed only as a formal 
power series and no closed formula is known for its coefficients. A list of analytic properties of conformal 
blocks which still require a deeper understanding contains the following problems: convergence issues of series 
defining conformal blocks, recurrence relations for coefficients, operator realization, integral 
representations, analytic continuations, classical limit.

From the point of view of applications the problem of calculation of the 
{\it classical limit} of the conformal blocks is 
currently the central issue concerning these functions.
In the classical limit the central charge
tends to infinity,
\begin{equation}\label{cTo}
c\;\longrightarrow\;\infty 
\end{equation}
and it is assumed, that
\begin{list}{}{\itemindent=2mm \parsep=0mm \itemsep=0mm \topsep=0mm}
\item[(a)] the weights are ``heavy'', i.e.,
$\frac{\Delta_{i}}{c}={\rm const.}\;$, 
$\frac{{\underline\Delta}_{p}}{c}={\rm const.}\;$ $\forall\;i,p$ or 
\item[(b)] the external weights are heavy (as above) and ``light'', i.e.,
$c^{-1}\Delta_{k}^{\rm light}\to 0\;$ for some
$k\in\lbrace 1,\ldots,n\rbrace$ or
\item[(c)] the weights are fixed, i.e.,
$\Delta_i={\rm const.}\;$,
${\underline\Delta}_p={\rm const.}\;$ $\forall\;i,p$.
\end{list}
For the standard parameterization of the central charge:
\begin{equation}\label{cVb}
c\;=\;1+6Q^2, \;\;\;\;\;\;\;\;\;\;
Q\;=\;b+\frac{1}{b}
\end{equation}
the limit (\ref{cTo}) corresponds to $b\to  0$ or $b\to\infty$.
In this parameterization the heavy weights are defined as follows:
\begin{eqnarray}\label{cl1}
(\Delta_i,{\underline\Delta}_p)=\frac{1}{b^2}(\delta_i,{\underline\delta}_p), 
\;&&\;
\delta_i,{\underline\delta}_p={\cal O}(b^0)\;\;\;{\rm if}\;\;\;b\to 0,
\\
\label{cl2}
(\Delta_i,{\underline\Delta}_p)=b^2(\delta_i,{\underline\delta}_p), 
\;&&\;
\delta_i,{\underline\delta}_p={\cal O}(b^0)\;\;\;{\rm if}\;\;\;b\to\infty.
\end{eqnarray}
In the first case the light conformal weights $\Delta_{k}^{\rm light}$ 
obey $\lim_{b\to 0}b^2\Delta^{\rm light}_{k}=0$, while in the second case we have,
$\lim_{b\to\infty}b^{-2}\Delta^{\rm light}_{k}=0$.
In the classical limit conformal blocks behave as follows.
\begin{list}{}{\itemindent=2mm \parsep=0mm \itemsep=0mm \topsep=0mm}
\item[(A)] 
If all the weights are heavy, then in the classical limit the blocks
exponentiate to functions known as Zamolodchikovs' \cite{ZZ5} 
{\it classical conformal blocks}:
\begin{equation}\label{cla1}\boxed{
{\cal F}\!\left(\lbrace\Delta_i\rbrace,
\lbrace{\underline\Delta_p}\rbrace, c\,|\,\cdot\,\right)
\;\stackrel{b\to\infty}{\sim}
\;{\rm e}^{b^2 f\left(\lbrace\delta_i\rbrace,\lbrace{\underline\delta_p}\rbrace|\,\cdot\,\right)}}
\end{equation}
(or analogously for $b\to 0$, where $b^2$ is replaced by $b^{-2}$ and relations (\ref{cl1}) are assumed).
\item[(B)] 
If the external weights are heavy and  light, 
$\lbrace \Delta_i \rbrace=\lbrace\Delta_l\rbrace\cup\lbrace\Delta_{k}^{\rm light}\rbrace$,
then in the classical limit the blocks
decompose into a product of the ``light contribution'' $\Psi_{\rm light}\left(\cdot\right)$  and the
exponent of the classical block:
\begin{equation}\label{cla2}\boxed{
{\cal F}\!\left(\lbrace\Delta_l\rbrace\cup\lbrace\Delta_{k}^{\rm light}\rbrace,
\lbrace{\underline\Delta_p}\rbrace, c\,|\,\cdot\,\right)
\;\stackrel{b\to \infty}{\sim}\Psi_{\rm light}\left(\cdot\right)\,
{\rm e}^{b^2 f\left(\lbrace\delta_l\rbrace,\lbrace{\underline\delta_p}\rbrace|\,\cdot\,\right)}}
\end{equation}
(or analogously for $b\to 0$, where $b^2$ is replaced by $b^{-2}$ and relations (\ref{cl1}) are assumed).
\item[(C)] 
If all the weights are fixed, then in the large central charge limit (\ref{cTo})
conformal blocks reduce to the so-called {\it global blocks}, 
i.e., contributions to the correlation functions
from representations of the $\mathfrak{sl}(2,\mathbb{C})$ algebra which is a global subalgebra
of the Virasoro algebra.
\end{list}
When commenting on the above statements, it should first be stressed that
the calculation of the global limit of conformal blocks 
(point (C) above) is incomparably simpler than a proof of (\ref{cla1}) and (\ref{cla2}).
For example, a four-point block on the sphere reduces in this limit 
to a hypergeometric function. Second,
although asymptotics (\ref{cla1}) and (\ref{cla2}) are confirmed by the classical limit of the 
quantum Liouville field theory (in the sense of saddle points of the correlation functions) 
and resulting from (\ref{cla1}), (\ref{cla2}) consequences, formulae (\ref{cla1}) and (\ref{cla2}) should be treated as 
conjectures rather than rigorous mathematical theorems. Third, it is worth noting that new results have 
recently been obtained in this field. For example, using the oscillator representation, it was possible 
to demonstrate an exponentiation of the spherical four-point block in the classical limit \cite{BDK}.
A mechanism of the factorization (\ref{cla2}) has been studied deeper in the case of  
the five-point spherical block containing four heavy external weights and one light external degenerate weight \cite{PPNPB}.
Finally, conjectures (\ref{cla1}) and (\ref{cla2}) have recently been extended to the so-called 
{\it irregular conformal blocks} \cite{LN,PP1,PP2,PP3}.

For a long time classical conformal blocks were known only 
to specialists dealing with the Liouville theory and its applications in 
the uniformization theory of Riemann surfaces (see~e.g.~\cite{ZZ5,HJP,HJ}).
In the last few years the classical limit of conformal blocks has appeared in novel mathematical contexts
and has been applied as a tool in the study of many
fundamental problems of contemporary theoretical physics.  
Some of these contexts and applications are cited below.

The novel mathematical contexts with classical blocks are for instance: 
the theory of the KdV equation \cite{WH}; the problem of isomonodromic deformations 
of the Fuchs equations \cite{Teschner:2017rve}, and newly discovered properties of 
the Painlev\'{e} VI  equation 
\cite{Litvinov:2013sxa} related to that problem.
Also monodromy problems for the Fuchs equations, implicitly unrelated to uniformization,
which have classical blocks as a solution, have been investigated at 
intensive rate (see~e.g.~\cite{Menotti:2014kra,Menotti:2016jut,Menotti:2018jsy,HK}).

In classical and quantum {\it physics of black holes}
the classical limit of conformal blocks is applied 
in the study of scattering problems of 
classical fields in geometries of certain black holes 
(see~e.g.~\cite{daCunha:2015fna,daCunha:2015ana,Novaes:2014lha,Amado:2017kao}), and
in the analysis of the famous information paradox
within 3-dimensional quantum gravity by means of the methods of the dual theory, namely CFT$_2$
(see~e.g.~\cite{Anous:2016kss,CHKL}).

The classical blocks have holographic counterparts via {\it AdS$_3$/CFT$_2$ correspondence}
(see~e.g.~\cite{Alkalaev:2016rjl,Alkalaev:2016ptm,Alkalaev:2015fbw,Alkalaev:2015lca,Alkalaev:2015wia,AlkPav}).
Moreover, the classical limit and the classical blocks appear here: 
in holographic calculations of entanglement entropy on the 2d CFT side (see the crucial paper  
\cite{Hartman:2013mia} and e.g.~\cite{Asplund:2014coa,Banerjee:2016qca,MH});
in a holographic interpretation of conformal bootstrap \cite{Ver};
in the study of holographic 2d CFTs (see~e.g.~\cite{HartmanIII,Chang:2015qfa,Chang:2016ftb}); 
in tests of eigenstate thermalization hypothesis (ETH) \cite{FW,FKW,FKap,LDL}.

Regardless of the holographic context, the classical limit of conformal blocks 
appears in the study of universal properties of the {\it entanglement entropy} in 2d CFT
(cf.~\cite{Hartman:2013mia} and e.g.~\cite{TakaK,Kusuki,Kusuki2}).
The entanglement entropy is a measure of how quantum information is stored in a 
quantum state. 
The entanglement entropy can be defined in quantum mechanics and in the quantum field theory including
quantum gravity. 

Black holes physics, three-dimensional quantum gravity, entanglement and holography are not all the current 
hot topics of theoretical physics in which the classical limit of conformal blocks is applied.
These techniques also appear in the research on {\it quantum chaos} (see the seminal paper
~\cite{RobStan}), the so-called
{\it topological strings} \cite{AKPT,AKPT2,AKP} and in the context of {\it matrix models} 
(cf.~\cite{BMT,Rim:2015tsa,Rim:2015aha}).
In addition, it turns out that classical conformal blocks have their counterparts in 
{\it supersymmetric ${\cal N}=2$ Yang--Mills theories} 
and in the theory of {\it quantum integrable systems}.

Indeed, in 2009 an amazing discovery known as the {\it AGT correspondence} \cite{AGT} was made. 
The AGT conjecture states that the Liouville field theory 
correlators on the Riemann surface $C_{g,n}$ with 
genus $g$ and $n$ punctures can be identified with the
partition functions of the class 
${\cal S}_{g,n}$ of four-dimensional ${\cal N}=2$ 
supersymmetric SU(2) quiver gauge
theories.\footnote{I.e., with the gauge group ${\rm SU}(2)^{\otimes 3g-3 + n}$,
$n$ flavors and $g$ loops in the quiver diagram representing the theory 
which corresponds to a pants decomposition of the surface $C_{g, n}$ 
consistent with the factorization of the correlator in the Liouville theory.}
A significant part of the AGT conjecture 
is an exact correspondence between the Virasoro
blocks on $C_{g,n}$ and the instanton sectors of the Nekrasov 
partition functions  \cite{N,NekraOkun} of the gauge theories
${\cal S}_{g,n}$. 
The AGT relations are provided when
an appropriate identification is established between the
parameters (see the diagram below). In particular, the conformal weights $\Delta_\alpha$ in the 
intermediate channels are expressed in terms of
the vacuum expectation values ${\bf a}$ of scalar fields in 
vector multiplets; the external weights $\Delta_\beta$
are functions of masses ${\bf m}$ of the matter fields; 
locations of the vertex operators correspond to complexified gauge couplings;
the background charge  $Q$ and therefore the central charge $c = 1 + 6Q^2$ are the functions of the 
$\Omega$  background parameters: $\epsilon_1, \epsilon_2$ or equivalently the quantum parameter of 
the Liouville theory is $b^2= \epsilon_2 / \epsilon_1$.

\bigskip
$$
\begin{tabular}{|c|}
  \hline
  ${\sf Liouville}\;{\sf correlation}\; {\sf function}$ \\
  $\left\langle \prod\limits_{a=1}^{n}V_{\beta_a}
  \right\rangle_{C_{g,n}}^{\sf LFT}$  \\
  ${\sf on}\; C_{g,n} - {\sf Riemann}\; {\sf surface}$ \\
  ${\sf with}\; {\sf genus}\; g \; {\sf and} \; n \; {\sf punctures}$ \\
  \hline
\end{tabular}
\;\;=\;\;
\begin{tabular}{|c|}
  \hline
  ${\sf Partition}\;{\sf function}\;Z_{{\cal S}(g,n)}$ \\
  ${\sf of}\;{\sf the}\;{\sf class}\;{\cal S}(g,n)\; {\sf of}\;4d$\\
  $(\Omega-{\sf deformed})\;{\cal N}=2\;{\sf SUSY}$\\
  ${\sf SU}(2)\;{\sf quiver}\; {\sf gauge}\;
  {\sf theories}$\\
  \hline
\end{tabular}
$$
$$
\hspace{-20pt}
\begin{tabular}{|c|}
  \hline
  ${\sf Virasoro}\;{\sf conformal}\;{\sf block}$ \\
  ${\cal F}_{1+6Q^2,\alpha}[\beta]({\sf Z})\;{\sf on} \; C_{g,n};$ \\
  $\alpha\equiv(\alpha_1, \ldots,\alpha_{3g-3+n}),$\\
   $\beta\equiv(\beta_1,\ldots,\beta_n),$  \\
  ${\sf Z}=(z_1, \ldots,z_{3g-3+n}),$\\
  $\Delta_{\alpha}=\alpha(Q-\alpha),\; Q=b+b^{-1}$\\
  \hline
\end{tabular}
=
\begin{tabular}{|c|}
  \hline
  ${\sf Nekrasov}\;{\sf instanton}\;{\sf partition}\;{\sf function}$ \\
  $\mathcal{Z}_{{\sf inst}}({\sf Z}, {\bf a}, {\bf m}; \epsilon_1, \epsilon_2);$ \\
  ${\bf a}=(a_1, \ldots, a_{3g-3+n}),$\\
  $ {\bf m}=(m_1, \ldots, m_n),$  \\
  ${\sf Z}=(z_i=\exp 2\pi \tau_i)_{i=1,\ldots,3g-3+n},$\\
  $(\epsilon_1+\epsilon_2)^2/(\epsilon_1\epsilon_2)=Q
\Longleftrightarrow\;\sqrt{\epsilon_2/\epsilon_1} = b$\\ 
  \hline
\end{tabular}
$$
\bigskip

The AGT correspondence works at the level of the quantum Liouville
field theory (LFT). At this point a question can be asked about
what happens if we proceed to the classical limit
of the Liouville theory. It turns out that the classical limit of 
LFT correlators corresponds to the known limit of partition 
functions of supersymmetric gauge theories --- the so-called 
{\it Nekrasov--Shatashvili limit} \cite{NS:2009}.
In particular, a consequence of that correspondence is that the classical
conformal blocks can be identified with the instanton sectors of the
{\it effective twisted superpotentials}. 
The latter quantities determine the low energy effective
dynamics of the two-dimensional gauge theories restricted to the
$\Omega$-background. The twisted superpotentials
play also a pivotal role in another duality, the  
{\it Bethe/Gauge  correspondence} \cite{NS:2009}. 
As a result of that duality,
the twisted superpotentials are identified with the {\it Yang--Yang functions}
of the corresponding quantum integrable systems (QIS).
These functions are potentials for the so-called {\it Bethe equations} 
which describe spectra of quantum integrable models.
Hence, combining the classical/Nekrasov--Shatashvili limit of the AGT duality and the Bethe/Gauge
correspondence, one gets thus the {\it triple correspondence} 
which links the 
classical blocks to the twisted superpotentials and then to the Yang--Yang (YY) functions.
Precisely, we have two options here.
First, let us note that the 2-particle QIS are just
quantum--mechanical systems. 
In this case classical Virasoro blocks and twisted superpotentials are determined by eigenvalues
of appropriate Schr\"{o}dinger operators,
\medskip
\begin{center}
\begin{tabular}{|r|l|c|}
\hline\hline
{\rm 2dCFT} & ${\rm 2d}\;{\cal N}=2\;$SU(2)\;SYM & {\rm 2-particle}\;{\rm QIS}\;\\ \hline\hline
classical Virasoro blocks &\;\;\;\;twisted superpotentials  & spectra of Schr\"{o}dinger operators\\ \hline
\end{tabular}
\end{center}
\medskip
Secondly, if we take the classical/Nekrasov--Shatashvili limit of the generalized AGT conjecture \cite{Wyllard:2009hg}
then the above-mentioned correspondence can be extended to the following 
\medskip
\begin{center}
\begin{tabular}{|r|l|c|}
\hline\hline
{\rm 2dCFT} & ${\rm 2d}\;{\cal N}=2$\;SU(N)\;{\rm SYM} & {${\rm N}$-particle}\;{\rm QIS}\;\\ \hline\hline
classical Toda blocks &\;\;\;\;  twisted superpotentials  & Yang--Yang functions              \\ \hline
\end{tabular}
\end{center}
\medskip
where on the gauge theory side there are supersymmetric ${\cal N}=2$ ${\rm SU(N)}$ theories.
Here, on the one hand, the twisted superpotentials 
${\cal W}^{\rm SU(N)}$ of the ${\rm SU(N)}$ gauge theories are 
the Yang--Yang functions for the N-particle QIS. On the other hand,  
${\cal W}^{\rm SU(N)}$ should correspond to classical Toda blocks.

So, taking into account applications, any new method of 
computing classical blocks 
and/or any new result concerning these functions 
can be of great use. 
In this paper we propose new finite expressions 
for certain multi-point classical Virasoro blocks on the sphere.
Our proposal is based on the Mironov--Morozov--Shakirov identities 
\cite{MMS} connecting power series representations of
Virasoro blocks and their Dotsenko--Fateev 
integral representations. The main results of this work are 
conjectured relations (\ref{fW}) and (\ref{ConM}) expressing 
classical Virasoro blocks in terms of saddle point values of 
Dotsenko--Fateev integrals. 

A particular motivation and inspiration for the present paper is
one  
yet more possible use of classical conformal blocks that was not mentioned above.
In the first paragraph of the introduction it is mentioned that 2d CFT has applications in 
condensed matter physics, for example in FQHE, where the Laughlin wave functions can be 
represented by certain conformal blocks calculated within a free field realization.
Interestingly,
an idea of representing quantum states by conformal 
blocks works great, when it is applied to characterize eigenstates 
of certain many-body quantum 
Hamiltonians describing pairing force interactions (Cooper pairs \cite{BCS}) 
or systems of interacting spins (Gaudin magnets \cite{Gaudin1,Gaudin2}). 
This concept has been used for the first time by 
Sierra in  the seminal paper \cite{Sierra:1999mp}, where he proposed 
a connection between 2d CFT and the exact solution, and integrability of
the so-called reduced BCS model defined by the Hamiltonian:
\begin{equation}\label{RBCSH}
\hat{\rm H}_{\rm rBCS}=
\sum\limits_{j,\sigma=\pm}\varepsilon_{j\sigma}c_{j\sigma}^{\dagger}c_{j\sigma}
-gd\sum\limits_{j,j'}c_{j+}^{\dagger}c_{j-}^{\dagger}c_{j'-}c_{j'+}.
\end{equation}
Let us remind that $c_{j\sigma}^{\dagger}$ and $c_{j\sigma}$ in eq.~(\ref{RBCSH}) are fermion 
creation and annihilation operators in the so-called time--reversed states 
$\ket{j,\pm}$ with energies $\varepsilon_j$,
$j=1,\ldots,\Omega$; $g$ stands for a dimensionless coupling constant and $d$ 
is a mean level spacing. 
The Hamiltonian (\ref{RBCSH}) is a simplified version of the
Bardeen--Cooper--Schrieffer (BCS) Hamiltonian, where all 
couplings have been set equal to a single one, namely $g$.
This {\it reduced} BCS model describes physics of ultra-small 
superconducting grains \cite{RBT,RBT2}, provided that a fixed number of fermion 
pairs is assumed (for a historical review of these investigations 
till 2001, see \cite{Sierra:2001cx}). 
This discovery caused a growing interest 
in the so-called Richardson's exact solution of the spectral problem
for $\hat{\rm H}_{\rm rBCS}$  (cf.~section \ref{RGmodels}) and  
led to a significant development of this field.
Today, the model defined by the Hamiltonian
(\ref{RBCSH}) is one of important representatives of quantum integrable systems 
from an entire class of the so-called Richardson--Gaudin models 
(see excellent introductions to the subject \cite{Dukelsky:2004re,Claeys:2018zwo}).\footnote{
Note, that a numerical analysis of this model for atomic nuclei with a small number 
of broken pairs was first carried out by A.~Pawlikowski and V.~Rybarska, see \cite{PawlikowskiRybarska}. 
Later, in series of papers (e.g., refs.~\cite{R1,RS,RS2}) R.W. Richardson (together with N.~Sherman) 
developed an analytical approach to this problem. Nowadays, this approach is recognized as the Richardson model.}

A connection of the above topics with the present work is as follows. 
As reported in \cite{Sierra:2001cx},
Gaudin and Richardson introduced the so-called ``electrostatic analogy'' 
(see section \ref{RGmodels}) and used it to solve the reduced BCS model in the thermodynamic limit when
the number of particles ($\Leftrightarrow$ fermion pairs) tends to infinity.\footnote{See also \cite{RSD}.}
Sierra embedded the electrostatic picture in conformal field theory \cite{Sierra:1999mp}.
Precisely, the author of \cite{Sierra:1999mp} has shown
that eigenfunctions and eigenvalues of the Hamiltonian (\ref{RBCSH})
are obtained in the saddle point limit from certain {\it deformed} Wess--Zumino--Witten (WZW)
conformal block in the Coulomb gas integral representation.
The saddle point limit used by Sierra is nothing but the classical limit of conformal blocks.
Thus, the results of \cite{Sierra:1999mp} should be closely related to the theory of classical blocks.
On the other hand, on ideas seen in \cite{Sierra:1999mp} and the 
additional technical tool --- 
the aforementioned MMS identities, are based derivations of the
main results of this work.
We believe that it is possible to work out similar techniques in case of WZW blocks.
This should lead to a developing new analytical tools for studying models of the type (\ref{RBCSH}).
Finally, it is also worth mentioning some known and less known correspondences (see Fig.1 and Fig.2) 
that would be applicable here, and 
their examination in this context could also pave a way for new analytical tools useful in the study of the model 
defined by (\ref{RBCSH}) and its possible generalizations.\footnote{ 
For instance from a combination of dualities depicted on first and second diagrams an intriguing 
question arises, does exist a dual  description of models of the type (\ref{RBCSH}) in supersymmetric 
gauge theory or at least whether one can use the techniques of SYM here.}
\begin{figure}[tbp]
\label{TI}
\begin{center}
\begin{tikzcd}[column sep=normal]
& {\rm rBCS} \arrow[dl, leftrightarrow, "{\br A}" description] 
\arrow[dr, leftrightarrow, "{\br C}" description] & \\
{\rm 2d}\;{\rm CFT} \arrow[rr, leftrightarrow, "{\br B}" description] & & {\rm matrix}\;{\rm models}
\end{tikzcd}
\end{center}
\caption{\it The arrow ${\br A}$ indicates the BCS/CFT correspondence \cite{Sierra:1999mp},
in which eigenfunctions and eigenvalues of the Hamiltonian (\ref{RBCSH})
are obtained in the saddle point limit from deformed WZW
conformal block in the Coulomb gas representation.
The arrow ${\br B}$ represents known formal agreement between the 
Dotsenko--Fateev \cite{DF,DF2} 
(Coulomb gas) integrals and Penner-type random matrix integrals \cite{Penner}.
The arrow ${\br C}$ stands for a link
between random matrices and integrable systems that can be named the Bethe/matrix models duality.
It manifests itself in that, that
for an appropriate choice of the potential, saddle point equations for the matrix model integral, i.e., 
equations for critical points of the matrix model action, are formally identical to the Bethe ansatz equations 
for some quantum integrable models. In particular, for certain Penner logarithmic 
potentials stationary points equations are 
identical to the Bethe ansatz equations of the reduced BCS model, see~e.g.~\cite{EyMar}.}
\end{figure}
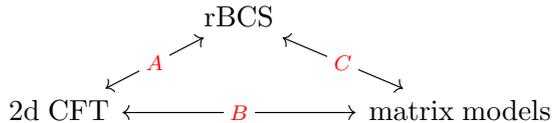
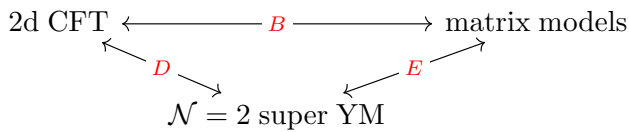
\begin{figure}[tbp]
\label{T2}
\begin{center}
\begin{tikzcd}[column sep=small]
{\rm 2d}\;{\rm CFT} \arrow[dr,leftrightarrow, "{\br D}" description] 
\arrow[rr, leftrightarrow, "{\br B}" description]
& & {\rm matrix}\;{\rm models} \arrow[dl, leftrightarrow, "{\br E}" description] \\
& {\mathcal{N}=2}\;{\rm super}\;{\rm YM} 
\end{tikzcd}
\end{center}
\caption{\it
Here the arrow ${\br D}$ is nothing but the AGT correspondence 
while the arrow ${\br E}$ represents the so-called matrix model 
version of the AGT duality \cite{DV}.}
\end{figure}

The paper is organized as follows.
Section \ref{Back} contains introductory material on the subject.
In particular, taking care of the self-sufficiency of this work, 
in subsection \ref{VirCFT} we collect 
material concerning 2d CFT on the sphere that can be found in 
textbooks.
Subsection \ref{Classical limit} gives a short introduction to the quantum Liouville theory on the sphere and  
its classical limit. Operator and geometric path integral approaches are discussed.
It is demonstrated in addition how the classical limit of Liouville correlation functions 
implies exponential beahviour (\ref{cla1}) of Virasoro blocks.
In section \ref{CVDF} we derive our main results, i.e., formulae 
(\ref{fW}) and (\ref{ConM}).\footnote{The simplest case of the 
formula (\ref{fW}) is numerically analyzed in appendix \ref{AppC}.}
Section \ref{RGmodels} is a review of the Richardson solution of the reduced BCS model, 
its relation to the Gaudin model, and conformal field theory realization 
of these models found by Sierra \cite{Sierra:1999mp}. At the end of section \ref{RGmodels}, 
we spell out our hypothesis about the possible relationship of 
the results obtained in \cite{Sierra:1999mp} with the classical limit of conformal blocks.
Section \ref{Con} provides a brief summary along with the issues that require further research.

\section{Background}
\label{Back}
\subsection{Virasoro conformal field theory}
\label{VirCFT}
In 1970 Polyakov \cite{Polyakov1970} conjectured that the physical system 
at the phase transition point should be described by
a conformally invariant field theory. The energy-momentum tensor in this theory is traceless. The system is 
described by a set of local fields $\left\lbrace\phi_{h}\right\rbrace$ 
with their scaling dimensions $h$ explicitly related to critical exponents.
Conformal invariance determines completely $2$- and $3$-point correlation functions.

To calculate the scaling dimensions, Polyakov proposed the bootstrap approach \cite{Polyakov1974}, 
based on the hypothesis of completeness of the operator algebra. This hypothesis says that the Wilson's 
expansion \cite{Wilson} of  the product of local operators (OPE --- operator
product expansion) is convergent. Due to OPE, any correlator of local fields can be expressed in several
ways through the $3$-point functions. The equivalence of different ways the correlation functions can be reduced
to the 3-point functions implies bootstrap equations that impose 
constraints on the scaling dimensions. In D-dimensional theory, when 
$\textrm{D}> 2$, the bootstrap equations had not been solved. 
The situation is different in two dimensions, where the 
conformal group is locally infinite dimensional.

Implementation of the bootstrap program in two-dimensional field theory is a famous work of Belavin, 
Polyakov and Zamolodchikov \cite{BPZ}, in which the concept of a conformal block emerged for the first time. 
BPZ gave a set of properties that define the non-lagrangian quantum field theory in two dimensions, where 
symmetries are local scalings of the metric.

In the BPZ theory, the conformal symmetry is generated by the holomorphic components of the
energy-momentum tensor: $T(z)$, $\bar T(\bar z)$, where modes $L_n$, $\bar L_n$ 
in the Laurent expansion form two copies of the Virasoro algebra,
\begin{equation}\label{Vir}
[L_n, L_m] = (n-m)L_{n+m}+\frac{c}{12}(n^3 - n)\delta_{n+m,0}
\end{equation}
with the same central charge $c$.
In the BPZ formulation of conformal field theory, a key role is played by the representation theory of the 
Virasoro algebra.

The so-called highest weight representation of the Virasoro algebra (\ref{Vir}) is the Verma module:
\begin{eqnarray}\label{basis}
{\cal V}_{c,\Delta} &=& \bigoplus_{n=0}^{\infty}{\cal V}_{c,\Delta}^{n},\nonumber
\\
{\cal V}_{c,\Delta}^{n} &=& \textrm{Span}\Big\lbrace
|\,\Delta^{n}_{I}\,\rangle=L_{-I}|\,{\Delta}\,\rangle 
:= L_{-i_{k}}\ldots L_{-i_{2}}L_{-i_{1}}|\,{\Delta}\,\rangle
\\
&&\hspace{20pt}:
I=(i_{k}\geq\ldots\geq i_{1}\geq 1)
\;\textrm{---}\; \textrm{partition} \;
\textrm{with}\; \textrm{length}\;n=i_k+\ldots+i_{1}=:|I|
\Big\rbrace,\nonumber
\end{eqnarray}
where $|\,{\Delta}\,\rangle$ is the highest weight vector, that obeys 
$L_0 |\,\Delta\,\rangle = \Delta |\,\Delta\,\rangle$ and $L_n |\,\Delta\,\rangle = 0$, $n>0$.
The eigenvalue $\Delta$ is called the conformal weight.
(The scaling dimension $h$ and conformal spin $s$ are given by 
$h=\Delta+\bar\Delta$ and $s=\Delta-\bar\Delta$, where $\Delta$ and $\bar\Delta$ are the highest weights of ``left'' and  ``right'' Virasoro representations respectively.)
The dimension of the space ${\cal V}_{c, \Delta}^{n}$ equals a number of partitions ${\sf p}(n)$ of the
positive integer $n$ called level or degree (${\sf p}(0)=1$). 
Subspaces ${\cal V}_{c, \Delta}^{n}$ are eigenspaces of $L_0$ to the eigenvalue $\Delta+n$.
A bilinear form (the Shapovalov form)  
$\langle\,\cdot\,|\,\cdot\,\rangle$ exists on ${\cal V}_{c, \Delta}^{n}$
(and therefore on ${\cal V}_{c, \Delta}$). The Shapovalov form is uniquely defined by the relations:
$\langle\,\Delta\,|\,\Delta\,\rangle=1$ and
$(L_n)^{\dagger}=L_{-n}$.
The Gram matrix $G_{c,\Delta}$ of the form $\langle\,\cdot\,|\,\cdot\,\rangle$
is block-diagonal in the basis $\left\lbrace|\,\Delta_I\,\rangle\right\rbrace$ with blocks of the form
\begin{equation}\label{Gram}
\Big[G_{c,\Delta}^{\,n}\Big]_{IJ}\;=\;\langle\,\Delta_{I}^{n}\,|\,\Delta_{J}^{n}\,\rangle.
\end{equation}

In any quantum field theory (QFT) the basic object of the study are correlation functions of quantum fields. 
In the operator formulation of 2d CFT developed  in \cite{BPZ}  
the $n$-point correlation functions are defined as vacuum expectation values of radially ordered products of 
local fields living on the Riemann sphere. 
As in the D-dimensional conformal field theory, in 2d CFT conformal invariance 
fixes completely the form of 2- and 3-point correlation functions. 
Moreover, in CFT$_2$ also a structure of the 4-point function is universal. So for example 
the $s$-channel 4-point function of primary fields decomposes 
into a sum of 4-point blocks multiplied by the structure constants,
\begin{eqnarray}\label{4-punktowa}
\langle\,\Delta_4\,|\phi_{\Delta_3}(1,1)\phi_{\Delta_2}(x, {\bar
x})|\,\Delta_1\,\rangle &=& \sum_{s}C(\Delta_{4},\Delta_{3},\underline{\Delta}_{s})
C(\underline{\Delta}_{s},\Delta_{2},\Delta_{1})\nonumber
\\
&&\times\,
\left|{\cal F}\!\left(\Delta_4,\ldots,\Delta_1,\underline{\Delta}_{s},c\,|\,x\,\right)\right|^{2}.
\end{eqnarray}
The representation (\ref{4-punktowa}) of
the 4-point function applies to spinless primary fields, i.e., to a
situation when $\Delta_i = \bar\Delta_i$. 
In the 4-point function (\ref{4-punktowa}), 
according to the so-called {\it field--state correspondence}, the {\it in}-state 
$|\,\Delta_1\,\rangle:=|\,\Delta_1\,\rangle\otimes|\,\bar\Delta_1\,\rangle$
and its dual, the {\it out}-state 
$\langle\,\Delta_4\,|:=\langle\,\Delta_4\,|\otimes\langle\,\bar\Delta_4\,|$
are tensor products of the highest weight vectors created by primary fields from the
${\rm SL}(2,{\mathbb C})$-invariant vacuum
$|\,0\,\rangle:=|\,0\,\rangle\otimes|\,0\,\rangle$:
$$
|\,\Delta\,\rangle\otimes|\,\bar\Delta\,\rangle
=\lim_{z,\bar z\to 0}\phi_{\Delta,\bar\Delta}(z,\bar z)|\,0\,\rangle,
\;\;\;\;\;\;\;\;\;\;\;\;\;\;
\langle\,\Delta\,|\otimes\langle\,\bar\Delta\,|
=\lim_{z,\bar z\to\infty}z^{2\Delta}{\bar z}^{2\bar\Delta}
\langle\, 0\,|\phi_{\Delta,\bar\Delta}(z,\bar z).
$$
In case of the continuous spectrum of the theory, 
the sum in (\ref{4-punktowa}) is replaced by an integral.
The structure constants $C(\Delta_{i},\Delta_{j},\Delta_{k})$ are nothing but the
3-point correlation functions calculated in the positions $\infty, 1, 0$ of
primary operators. These quantities are model dependent and encode dynamics of concrete 2d CFTs.
The 4-point block ${\cal F}\!\left(\Delta_4,\ldots,\Delta_1,\underline{\Delta}_{s},c\,|\,x\,\right)$ 
represents contribution to the 4-point correlation function from the representation 
${\cal V}_{c,\underline{\Delta}_s}$ which propagates in the intermediate channel.
The exact form of the generic 4-point block is unknown, however this function is available for 
calculations as a formal power series with coefficients completely determined by the conformal symmetry.

Indeed, in 2d CFT
the operator product expansion of primary fields can be written as
\begin{equation}
\phi_{\Delta_1,\bar \Delta_1}(x,\bar {x})\phi_{\Delta_2,\bar \Delta_2}(0,0)=\sum_{s}
C_{12s}
\,x^{\underline{\Delta}_{s}-\Delta_{1}-\Delta_{2}}
\overline{x}^{\,\,\underline{\bar\Delta}_{s}-\bar{\Delta}_{1}-\bar{\Delta}_{2}}
\Psi_{12s}(x,\bar x),
\end{equation}
where for each $s$ the descendent field
$\Psi_{12s}(x,\bar x)$ is uniquely determined by the conformal invariance.
Acting on the vacuum $|\,0\, \rangle $ it generates a state 
$\Psi_{12s}(x,\bar x)
|\,0\,\rangle =|\,\psi_{12s}(x)\,\rangle\otimes
|\,\bar \psi_{12s}(\bar{x})\,\rangle$ in
the tensor product
${\cal V}_{c,\underline{\Delta}_s}\otimes {\cal V}_{c,\underline{\bar\Delta}_s}$
of the Verma modules with the
the highest weights $\underline{\Delta}_{s}$, and $\underline{\bar\Delta}_s$, respectively.
The $x$ dependence of each component is uniquely determined by the conformal invariance.
In the left sector one has
\begin{equation}\label{psi12s}
|\,\psi_{12s}(x)\,\rangle \;=\;\left|\,\underline{\Delta}_s\,\right\rangle + \sum\limits_{n=1}^\infty x^n
\,\left|\,\beta^{\,n}_{ c,\underline{\Delta}_{s}}\!\left[\,^{\Delta_2}_{\Delta_1}\right]\,\right\rangle,
\end{equation}
where $\left|\,\underline{\Delta}_s\,\right\rangle$ is the highest weight vector in
${\cal V}_{c,\underline{\Delta}_{s}}$ and
$\left|\,\beta^{\,n}_{ c,\underline{\Delta}_{s}}\!\left[\,^{\Delta_2}_{\Delta_1}\right]\,\right\rangle
\in {\cal V}_{c,\underline{\Delta}_{s}}^n\subset {\cal V}_{c,\underline{\Delta}_{s}}$.

The conformal Ward identity for the
3-point function implies  the equations
$$
L_i\left|\,\beta^{\,n}_{ c,\underline{\Delta}_{s}}\!\left[\,^{\Delta_2}_{\Delta_1}\right]\,\right\rangle
= (\underline{\Delta}_{s} + i\Delta_1 -\Delta_2 + n-i)
\left|\,\beta^{\,n-i}_{c,\underline{\Delta}_{s}}\!\left[\,^{\Delta_2}_{\Delta_1}\right]\,\right\rangle
$$
which in the basis (\ref{basis}) take the form
\begin{equation}
\label{betas}
\sum\limits_{|J|=n} \Big[ G_{c,\underline{\Delta}_{s}}^{\,n}\Big]_{IJ}
\beta^{\,n}_{c,\underline{\Delta}_{s}}\!\left[\,^{\Delta_2}_{\Delta_1}\right]^J
\;=\;
\gamma_{\underline{\Delta}_{s}}\!\left[\,^{\Delta_2}_{\Delta_1}\right]_I\,,
\end{equation}
where 
\begin{eqnarray}
\label{gamma}
\gamma_{\underline{\Delta}_{s}}\!\left[\,^{\Delta_2}_{\Delta_1}\right]_I\!\! &=&\!\!
(\underline{\Delta}_{s} +i_k\Delta_1 -\Delta_2 + i_{k-1} \!+\!\dots \!+\!i_1)\times \dots\\
\nonumber
&\dots &\times(\underline{\Delta}_{s} +i_2\Delta_1 -\Delta_2 +i_1) (\underline{\Delta}_{s} +i_1\Delta_1 -\Delta_2 ) \ .
\end{eqnarray}
For all  values of the variables $c$ and $\underline{\Delta}_{s}$ for which the Gram matrices are invertible
the equations (\ref{betas}) admit  unique solutions
$$
\beta^{\,n}_{c,\underline{\Delta}_{s}}\!\left[\,^{\Delta_2}_{\Delta_1}\right]^I
=
\sum\limits_{|J|=n} \Big[ G_{c,\underline{\Delta}_{s}}^{\,n}\Big]^{IJ}
\gamma_{\underline{\Delta}_{s}}\!\left[\,^{\Delta_2}_{\Delta_1}\right]_J\,.
$$
In this range the 4-point conformal block defined as a product of vectors (\ref{psi12s}) 
has for fixed positions $\infty,1,x,0$ a formal power series representation
\cite{BPZ}:
\begin{eqnarray}
\label{block}
{\cal F}\!\left(\Delta_4,\ldots,\Delta_1,\underline{\Delta}_{s},c\,|\,x\,\right)
&=&
x^{\underline{\Delta}_{s}-\Delta_{2}-\Delta_{1}}\left( 1 +
\sum_{n=1}^\infty x^{n}
{\sf B}_{n}\!\left(\Delta_4,\ldots,\Delta_1,\underline{\Delta}_{s},c\right)\right),
\\
\label{blockBn}
{\sf B}_{n}\!\left(\Delta_4,\ldots,\Delta_1,\underline{\Delta}_{s},c\right)
&=&
\sum\limits_{|I|=|J|=n}
\gamma_{\underline{\Delta}_{s}}\!\left[\,^{\Delta_3}_{\Delta_4}\right]_I
\Big[G_{c,\underline{\Delta}_{s}}^{\,n}\Big]^{IJ}
\gamma_{\underline{\Delta}_{s}}\!\left[\,^{\Delta_2}_{\Delta_1}\right]_J.
\end{eqnarray}

The BPZ theory was generalized by Moore and Seiberg (MS) 
\cite{MS1, MS2} to the so-called extended conformal algebras or vertex algebras. Examples of the latter are  
Kac-Moody affine algebras, W-algebra \cite{FatZa}, symmetry algebras in theories of
para-fermions \cite{FatZa2}.
Moore and Seiberg gave axioms of the so-called rational conformational field theories (RCFT) where the 
Hilbert space of states is a representation space of the vertex (chiral) algebra
$ {\cal A} $. The vertex algebra ${\cal A} $ is a subalgebra of holomorphic fields of
the OPE algebra of local operators. 
In any 2d CFT there exist at least two chiral fields, i.e., the identity operator and
its descendant --- the holomorphic component of the energy-momentum tensor.
Therefore, each chiral algebra ${\cal A}$ contains as a subalgebra the Virasoro
algebra.

In the Moore--Seiberg formalism, the BPZ ``physical'' primary fields, $\phi_{\Delta,\bar\Delta}(z,\bar z)$, 
are built out of more fundamental objects --- the so-called chiral vertex operators (CVOs),
$\phi_{\Delta,\bar\Delta}(z,\bar z)\propto V_{\Delta}(z)\otimes V_{\bar\Delta}(\bar z)$, 
acting between representations of the vertex algebra.
Conformal blocks in such formalism are defined as matrix elements of appropriate compositions of CVOs.
If ${\cal A}$ is the Virasoro algebra, then the MS
$4$-point block in the $s$-channel it is the BPZ block (\ref{block}) with coefficients of the form:\footnote{
To distinguish weights in intermediate channels from external conformal weights, we mark the first ones by underlining. 
In the 4-point case, there is only one  intermediate weight, so we will omit this underline sometimes.}
\begin{equation}
\label{blockCeef}
{\sf B}_{n}\!\left(\Delta_4,\ldots,\Delta_1,\Delta,c\right) \;=
\sum\limits_{n=|I|=|J|}
\left\langle\,\Delta_4\,| V_{\Delta_3}(1) |\,\Delta_{I}^{n}\,\right\rangle
\;\Big[G_{c,\Delta}^{n}\Big]^{IJ}
\;\left\langle\,\Delta_{J}^{n}\,| V_{\Delta_2}(1) |\,\Delta_1\,\right\rangle.
\end{equation}
Just like before, in eq.~(\ref{blockCeef}) the matrix $\Big[G_{c,\Delta}^{n}\Big]^{IJ}$ 
is an inverse of the Gram matrix $(\ref{Gram})$, while
$$
V_{\Delta_j}(z)=V\left(|\,\Delta_j\,\rangle | z\right)=
V_{\Delta_k,\Delta_i}^{\;\Delta_j}\left(|\,\Delta_j\,\rangle | z\right):
{\cal V}_{\Delta_i}\longrightarrow {\cal V}_{\Delta_k}
$$
are the primary CVOs that act between Verma modules and are defined by
(i) commutation relations with Virasoro generators,
$$
\left[L_n , V_{\Delta}(z)\right] = z^{n}\left(z
\frac{{\rm d}}{{\rm d}z} + (n+1)\Delta\right)V_{\Delta}(z);
$$ 
(ii) normalization 
$\left\langle\,\Delta_3\,| V_{\Delta_2}(z) |\,\Delta_{1}\,\right\rangle=z^{\Delta_3-\Delta_2-\Delta_1}$.
By employing an oscillator representation of the Virasoro algebra the primary CVOs can 
be realized in the Fock space as ordered exponentials of a free scalar field times the so-called screening charges (see e.g. 
\cite{DiFMS} and section \ref{CVDF}).

In the MS formalism, as in the ``physical'' conformal field theory constructed by BPZ, there is an isomorphism 
between vectors from the representation space of the vertex algebra 
and CVOs, i.e., 
$\lim_{z\to 0}V\left(|\,\zeta_j\,\rangle | z\right) |\,0\,\rangle=|\,\zeta_j\,\rangle$. 
If the vertex algebra is the Virasoro algebra then $|\,\zeta_j\,\rangle\in {\cal V}_{\Delta_j}$.

\subsection{Classical limit}
\label{Classical limit}
The quantum Liouville theory on the Riemann sphere
$\hat{\mathbb{C}} = \mathbb{C}\cup\lbrace\infty\rbrace $ in the {\it operator approach} \cite{ZZ5} 
it is the BPZ conformal field theory with the central charge  
$c_{\sf L}=1+6Q^2$, where $Q=b+\frac{1}{b}$ and the continuous spectrum of conformal weights
$\Delta_\alpha=\bar\Delta_\alpha=\alpha(Q-\alpha)$, 
$\alpha\in\mathbb{S}=\frac{1}{2}Q+i\mathbb{R}_{\geq 0}$.
The Hilbert space of states is a direct integral,
$$
{\cal H}_{\sf L}=\int\limits_{\mathbb{S}}\textrm{d}\alpha \ {\cal
V}_{\alpha} \otimes\bar{\cal V}_{\alpha}.
$$  
It is assumed that in
${\cal H}_{\sf L}$ exists the SL($2, \mathbb{C}$)-invariant vacuum ``ket''
$|\,0\,\rangle$ and the SL($2, \mathbb{C}$)-invariant, ``charged''
vacuum ``bra'' $\langle\,Q\,|$.
The Liouville vertex operators $\textsf{V}_{\alpha}(z,\bar z)$ are  
primary fields in the theory. These operators are quantum analogues
of classical exponents
$\textrm{e}^{2\alpha\phi}$, where $\varphi = 2b\phi$ is the
classical Liouville field. It is assumed that $\textsf{V}_{\alpha}(z,
\bar z)$ create the highest weight states:
\begin{eqnarray*}
|\,\alpha\,\rangle &=& \lim\limits_{z, \bar z\to 0}
\textsf{V}_{\alpha}(z, \bar z)|\,0\,\rangle,\\
\langle\,Q - \alpha\,|&=& \lim\limits_{z, \bar z\to\infty}
|z|^{\Delta_\alpha} \langle\,Q\,|\textsf{V}_{\alpha}(z, \bar z),
\;\;\;\;\;\;\;\;\;\; \bar\alpha = Q-\alpha.
\end{eqnarray*}
The dynamics of the model is encoded in the DOZZ 3-point function 
\cite{ZZ5,DO}:
\begin{eqnarray*}
\label{DOZZ}
&&C(\alpha_1, \alpha_2, \alpha_3)\;=\; \left[
\pi\mu\gamma(b^2)b^{2-2b^2}\right]^{(Q-\alpha_1 - \alpha_2 -
\alpha_3)/b}
\times\nonumber
\\[5pt]
&&\hspace{-50pt}\frac{\Upsilon_{0}\Upsilon(2\alpha_1)\Upsilon(2\alpha_2)\Upsilon(2\alpha_3)}
{\Upsilon(\sum \alpha_i - Q)\Upsilon(Q+\alpha_1 - \alpha_2 -
\alpha_3) \Upsilon(\alpha_1 + \alpha_2 - \alpha_3)\Upsilon(\alpha_1
- \alpha_2 + \alpha_3)},
\end{eqnarray*}
$$
\gamma(x)\equiv\frac{\Gamma(x)}{\Gamma(1-x)},
\;\;\;\;\;\;\;\;\;\;
\Upsilon_{0}=\frac{d\Upsilon(x)}{d x}\Big|_{x=0}
$$
  
In the {\it geometric path integral approach}
the correlators of the Liouville theory are expressed in terms of path integrals
over the conformal class of Riemannian metrics
with prescribed singularities at the punctures. In particular, in the case
of the quantum Liouville theory on the sphere the central objects
in the geometric approach are
\begin{enumerate}
\item
the ``partition functions'' on $C_{0,n}$:
\begin{equation}
\label{partition}
\left\langle\; C_{0,n} \;\right\rangle =
\int\limits_{\cal M}{\cal D}\varphi\;{\rm e}^{-Q^2 S_{\sf L}[\varphi]},
\end{equation}
where $\cal M$ is the space of conformal factors appearing in the
metrics $\textrm{e}^{\varphi(z,\bar z)} {\rm d}z {\rm d}\bar z$ on the $n$-punctured Riemann sphere 
$C_{0,n} =\hat{\mathbb{C}}\setminus\lbrace z_1,\ldots, z_{n-1},\infty\rbrace$
with either the asymptotic behavior of elliptic type,
\begin{equation}
\label{elliptic} \varphi(z,\bar z) = \left\{
\begin{array}{lll}
-2\left(1-\xi_j \right)\log | z- z_j |  + O(1) & \textrm{as}  &
z\to z_j,
\hskip 5mm j = 1,\ldots,n-1,\\
-2\left(1+\xi_n \right)\log | z| + O(1) & \textrm{as} & z\to
\infty,
\end{array}
\right.
\end{equation} 
or parabolic type, 
\begin{equation}
\label{elliptic} \varphi(z,\bar z) = \left\{
\begin{array}{lll}
-2\log |z- z_j |  -2\log \left|\log |z- z_j |\right| + O(1) & {\rm as } & z\to z_j, \\
-2\log |z| - 2\log \left|\log |z|\right| + O(1) & {\rm as } & z\to
\infty.
\end{array}
\right.
\end{equation}
and 
$S_{\sf L}[\varphi]= 
\frac{1}{4\pi}
\lim_{\epsilon\to 0}
S_{\sf L}^\epsilon[\varphi]$ is the regularized action with $S_{\sf L}^\epsilon[\varphi]$ 
given by 
\begin{eqnarray}
\label{actionreg} 
S_{\sf L}^\epsilon[\varphi] & = &
\int\limits_{\Omega_\epsilon}\!\textrm{d}^2z
\left[\left|\partial_z\varphi\right|^2 +
\mu\textrm{e}^{\varphi}\right] +
\sum\limits_{j=1}^{n-1}\left(1-\xi_j\right)
\hspace{-4mm}\int\limits_{|z-z_j|=\epsilon}\hspace{-4mm}|\textrm{d}z|\
\kappa_z \varphi +\left(1+\xi_n\right)
\hspace{-2mm}\int\limits_{|z|=\frac{1}{\epsilon}}\hspace{-2mm}|\textrm{d}z|\
\kappa_z \varphi \nonumber
\\
&& - 2\pi\sum\limits_{j=1}^{n-1}\left(1-\xi_j\right)^2\log\epsilon -
2\pi\left(1+\xi_n\right)^2\log\epsilon,
\end{eqnarray}
and \( \Omega_\epsilon = {\mathbb
C}\setminus\left\{\left(\bigcup_{j=1}^n |z-z_j|< \epsilon\right)
\cup \left(|z|>\frac{1}{\epsilon}\right)\right\}\);
\item
the correlation functions of the energy-momentum tensor:
\begin{eqnarray}
\label{emom}
\left\langle
\widehat{T}(u_1)\ldots \widehat{T}(u_k) \widehat{\bar T}(\bar w_1)
\ldots \widehat{\bar T}(\bar w_l)\, C_{0,n}
\right\rangle  &=&\nonumber
\\
&&\hspace{-100pt}=\;
\int\limits_{\cal M}\!\!{\cal D}\varphi\;{\rm e}^{-Q^2 S_{\sf L}[\varphi]}\;
\widehat{T}(u_1)\ldots \widehat{T}(u_k) \widehat{\bar T}(\bar w_1)
\ldots \widehat{\bar T}(\bar w_l)
\end{eqnarray}
with
\begin{equation}
\label{emom:def}
\widehat{T}(u) = Q^2\left[
-\frac{1}{4}\left(\partial_u\varphi(u,\bar u)\right)^2 + \frac{1}{2}\,\partial^2_u\varphi(u,\bar u)
\right].
\end{equation}
\end{enumerate}
One can check by perturbative calculations of the correlators (\ref{emom})
that the central charge reads $c_{\sf L} = 1 + 6Q^2$.
The transformation properties of (\ref{partition}) with respect to global conformal
transformations show that the punctures behave as primary fields with dimensions
\begin{equation}
\label{Delta}
\Delta_j =
\bar{\Delta}_j = \frac{Q^2}{4}\left(1-\xi_j^2\right).
\end{equation}
For fixed $\xi_j$, the dimensions (\ref{Delta}) scale like $Q^2$ and
the punctures correspond to {\it heavy} fields
of the operator approach \cite{ZZ5}.

In the classical limit $b\to\infty$ or $b\to 0$ with all classical weights
$\delta_i   = \frac{1}{4}(1-\xi_j^2)$
fixed,  we expect the path integral to be
dominated by the classical action $S^{\,\sf cl}_{\sf L}(\delta_i\,;\,z_i)$,
\begin{equation}
\label{asymptotic:X}
\left\langle \; C_{0,n} \; \right\rangle
\stackrel{b\to 0}{\sim}
{\rm e}^{-\frac{1}{b^2}S^{\,\sf cl}_{\sf L}(\delta_i\,;\,z_i)}.
\end{equation}
$S^{\,\sf cl}_{\sf L}(\delta_i\,;\,z_i)$
in eq.~(\ref{asymptotic:X}) denotes the regularized
functional $S_{\sf L}[\,\cdot\,]$ of eq.~(\ref{actionreg}) evaluated at the
classical solution $\varphi$ of the Liouville equation  with the elliptic or parabolic asymptotics.

The relationship between the path integral geometric and
the operator formulation of the quantum  Liouville theory on the sphere
is not fully understood 
\cite{Takhtajan5, MenoTon1, MenoTon2, Meno1}.
It is supposed that the classical limit of the DOZZ theory exists
and it is correctly described by the classical
Liouville action. This supposition is confirmed, for example, by the existence of 
the classical limit of the DOZZ 3-point function 
\cite{ZZ5} (see below).
The correspondence between the path integral geometric  and
the operator formulations of the quantum Liouville theory allows to postulate the classical asymptotics of 
conformal blocks.
In the next subsections below we present the derivation of the classical limit of the DOZZ 3-point function
and the classical asymptotic of the 4-point block on the 
sphere obtained by Zamolodchikovs \cite{ZZ5}.

\subsubsection*{Classical limit of the DOZZ 3-point function}
In \cite{Hadasz:2003he} it has been found that for the hyperbolic spectrum:
\begin{equation}\label{alphas}
\alpha_j \;=\; \frac{Q}{2}(1+i\lambda_j)
\;\stackrel{b\to 0}{\sim}\;
\frac{1}{2b}(1+i\lambda_j),
\;\;\;\;\;\; \lambda_j \in \mathbb{R},
\;\;\;\;\;
j=1,2,3
\end{equation}
the DOZZ three-point function in the limit $b\to 0$ behaves as follows
\begin{eqnarray}\label{DOZZlimit}
C\!\left(\alpha_1, \alpha_2, \alpha_3\right) &\sim & \exp\left\lbrace -\frac{1}{b^2}
\left[\sum\limits_{\sigma_1, \sigma_2 = \pm}
F\left( \frac{1+i\lambda_1}{2} + \sigma_1 \frac{i\lambda_2}{2} + \sigma_2 \frac{i\lambda_3}{2}\right)\right.\right.
\nonumber\\
&&+\sum\limits_{j=1}^{3}\left(H(i\lambda_j)+\frac{1}{2}\pi |\lambda_j|\right)+\frac{1}{2}\log(\pi\mu b^2)
\nonumber\\
&&\left.\left.-i\sum\limits_{j=1}^{3}\lambda_j \left( 1-\log|\lambda_j|+\frac{1}{2}\log(\pi\mu b^2)\right)
+ {\rm const.}
\right]
\right\rbrace,
\end{eqnarray}
where
$$
F(x)\;=\;\int\limits_{\frac{1}{2}}^{x}dy \log\frac{\Gamma(y)}{\Gamma(1-y)},
\;\;\;\;\;\;\;\;\;\;\;
H(x)\;=\;\int\limits_{0}^{x}dy \log\frac{\Gamma(-y)}{\Gamma(y)}.
$$
At this point, a few comments are in order.
The expression in the square brackets should correspond to the known expression
for the classical Liouville action $S_{\rm L}^{(3)}[\varphi]$
on the sphere with three hyperbolic singularities (holes).
Such classical action has been constructed in \cite{Hadasz:2003he}.
The construction of $S_{\rm L}^{(3)}[\varphi]$
relies on a solution of a certain monodrommy problem
for the Fuchsian differential equation:
$$
\frac{d^2 \Phi}{dz^2} + \sum_{k=1}^{n}\left[\frac{\delta_k}{(z - z_k)^2}+
\frac{c_k}{z - z_k}\right]\Phi \;=\; 0
$$
with hyperbolic singularities ($\delta_k$'s are hyperbolic).
In this way one can find the form of the $n$-point classical action
$S_{\rm L}^{(n)}[\varphi]$
up to of at most $n-3$ undetermined constants
$c_k$. $S_{\rm L}^{(n)}[\varphi]$
satisfies Polyakov's formula:
\begin{equation}\label{Polyakov}
\frac{\partial}{\partial z_j}S_{\rm L}^{(n)}[\varphi] \;=\; -c_j.
\end{equation}
In the case when $n=3$ the {\it Fuchsian accessory parameters} $c_k$ are known
and the classical action can be determined from eq. (\ref{Polyakov}).
For the standard locations of singularities $z_1 = 0$, $z_2 = 1$, $z_3 = \infty$
this yields \cite{Hadasz:2003he}:
\begin{eqnarray}\label{cl}
Q^2 S_{\rm L}^{(3)}[\varphi] &=&
Q^2 \left[ \sum\limits_{\sigma_1, \sigma_2 = \pm}
F\left( \frac{1+i\lambda_1}{2} + \sigma_1 \frac{i\lambda_2}{2} + \sigma_2 \frac{i\lambda_3}{2}\right)\right.
\nonumber\\
&+&\left.\sum\limits_{j=1}^{3}\left(H(i\lambda_j)+\frac{1}{2}\pi |\lambda_j|\right)+\frac{1}{2}\log(\pi\mu b^2)
+ \frac{1}{Q^2}\,{\rm const.}
\right],
\end{eqnarray}
where the constant on the r.h.s. is independent of $z_j$, $\lambda_j$ and $\pi\mu b^2$.
Comparing (\ref{DOZZlimit}) and (\ref{cl}) we see that the classical limit
of the DOZZ structure constant differs from the classical three-point action
by an additional imaginary term. As has been observed in \cite{Hadasz:2003he}
this inconsistency occurs due to the fact that the classical Liouville action
is by construction symmetric with respect to the reflection
$\alpha\to Q-\alpha$, $(\lambda\to -\lambda)$
whereas the DOZZ three-point function is not.
Under this reflection the DOZZ three-point function changes according to the formula
\cite{ZZ5}:
$$
C\left(Q-\alpha_1, \alpha_2, \alpha_3\right)\;=\;
{\cal S}(i\alpha_1 - iQ/2)C\left(\alpha_1, \alpha_2, \alpha_3\right),
$$
where
\begin{equation}
\label{reflection}
{\cal S}(x)\;=\;-\left(\pi\mu\gamma(b^2)\right)^{-2ix/b}
\frac{\Gamma(1+2ix/b)\Gamma(1+2ixb)}{\Gamma(1-2ix/b)\Gamma(1-2ixb)}
\end{equation}
is the so-called {\it reflection amplitude} \cite{ZZ5}.
The discrepancy between (\ref{DOZZlimit}) and (\ref{cl}) can
be overcome if we consider the \textit{symmetric} three-point function
$\tilde C\left(\alpha_1, \alpha_2, \alpha_3\right)$ \cite{Hadasz:2003he}:
\begin{eqnarray}\label{sDOZZ}
\tilde C\left(\alpha_1, \alpha_2, \alpha_3\right) &\equiv&
\left[\prod\limits_{j=1}^{3}\sqrt{{\cal S}(i\alpha_j - iQ/2)}\right]
C\left(\alpha_1, \alpha_2, \alpha_3\right)
\end{eqnarray}
instead of $C\left(\alpha_1, \alpha_2, \alpha_3\right)$.
Indeed, taking into account the classical limit of the
reflection amplitude for $\lambda\in \mathbb{R}$:
$$
\log {\cal S}\!\left(-\frac{\lambda}{2b}\right)\;\sim\; \frac{2i}{b^2}\lambda
\left(1-\log|\lambda|+\frac{1}{2}\log(\pi\mu b^2)\right)
$$
one can easily verify that the symmetric three-point function in the
limit $b\to 0$ behaves as follows \cite{Hadasz:2003he}
\begin{eqnarray}\label{symmDOZZ}
\tilde C\!\left(\alpha_1, \alpha_2, \alpha_3\right) &\sim &
\exp\left\lbrace -\frac{1}{b^2}
\left[\sum\limits_{\sigma_1, \sigma_2 = \pm}
F\left( \frac{1+i\lambda_1}{2} + \sigma_1 \frac{i\lambda_2}{2} + \sigma_2 \frac{i\lambda_3}{2}\right)\right.\right.
\nonumber\\
&&\left.\left.+\sum\limits_{j=1}^{3}\left(H(i\lambda_j)+\frac{1}{2}\pi |\lambda_j|\right)+\frac{1}{2}\log(\pi\mu b^2)
+{\rm const.}\right]
\right\rbrace\nonumber
\\
&=&
\exp\left\lbrace-\frac{1}{b^2}\,S_{\rm L}^{(3)}(\lambda_1, \lambda_2, \lambda_3)\right\rbrace,
\end{eqnarray}
where $\alpha_j$, $j=1,2,3$ are given by (\ref{alphas}).

\subsubsection*{Classical limit of the DOZZ 4-point function}
The partition function (\ref{partition}) corresponds in the
operator formulation to the correlation function of the primary
fields
$\textsf{V}_{\alpha_j}(z_j,\bar z_j)$:
\begin{equation}
\label{relation} \left\langle\, C_{0,n}\, \right\rangle  =
\Big\langle\textsf{V}_{\alpha_n}(\infty,\infty)\ldots
\textsf{V}_{\alpha_1}(z_1,\bar z_1) \Big\rangle,
\end{equation}
where
$\Delta_j = \alpha_j(Q-\alpha_j)$ and $\alpha_j =
\frac{1}{2}Q\left(1 - \xi_j\right)$.
Consider the case $n=4$.
Let us remember that the DOZZ 4-point function for the standard locations
$z_4 = \infty, z_3 = 1, z_2 = x, z_1 = 0$
is expressed as an integral over the continuous spectrum:
\begin{eqnarray}
\label{four:point:} && \hspace*{-1.5cm} \Big\langle
\textsf{V}_{\alpha_4}(\infty,\infty)\textsf{V}_{\alpha_3}(1,1)
\textsf{V}_{\alpha_2}(x,\bar x)\textsf{V}_{\alpha_1}(0, 0)
\Big\rangle =
\\
\nonumber && \int\limits_{\frac{Q}{2} + i{\mathbb
R}_{\geq 0}}\!\!\!\!\!\!\!\textrm{d}\alpha\; 
C(\alpha_4,\alpha_3,\alpha)C(Q-\alpha,\alpha_2,\alpha_1)
\left| 
{\cal F}\!\left(\Delta_4,\ldots,\Delta_1,\Delta,1+6Q^2\,|\,x\,\right)\right|^2.
\end{eqnarray}
Let $\mbox{\bf P}_{\Delta,\Delta}$ stand for a projection operator onto single
conformal family of the primary field with the conformal weights 
$\Delta=\bar\Delta$. 
The 4-point correlation function (\ref{four:point:}) with the $\mbox{\bf P}_{\Delta,\Delta}$
insertion factorizes into the product of the holomorphic and
anti-holomorphic factors:
\begin{eqnarray}
\label{c4} 
 \Big\langle
\textsf{V}_4(\infty,\infty)\textsf{V}_3(1,1)\mbox{\bf
P}_{\Delta,\Delta} \textsf{V}_2(x,\bar x)\textsf{V}_1(0,0)
\Big\rangle &=& C(\alpha_4,\alpha_3,\alpha)
C(Q-\alpha,\alpha_2,\alpha_1)\nonumber
\\
&&\times{\cal F}(x)\bar{\cal F}(\bar x).
\end{eqnarray}
Assuming the path integral representation of the left hand side, one
should expect in the limit $b\to 0$, with all the weights being heavy,
i.e.,~$\Delta,\Delta_i \sim \frac{1}{b^2}\cdot{\rm const.}$,
the  following asymptotic behavior:
\begin{equation}
\label{a4} \Big\langle
\textsf{V}_4(\infty,\infty)\textsf{V}_3(1,1)\mbox{\bf
P}_{\Delta,\Delta} \textsf{V}_2(x,\bar x)\textsf{V}_1(0,0)
 \Big\rangle \sim
\textrm{e}^{-\frac{1}{b^2} S^{\rm cl}(\delta_i,x;\delta) }.
\end{equation}
However, the $b\to 0$ limit of the DOZZ
coupling constants gives as a result
\begin{equation}
\label{asymptotC} C(\alpha_4,\alpha_3,\alpha)C(Q-\alpha,\alpha_2,\alpha_1)
 \sim
\textrm{e}^{-\frac{1}{b^2}\left( S^{\rm cl}(\delta_4,\delta_3,\delta) +
S^{\rm cl}(\delta,\delta_2,\delta_1) \right)}.
\end{equation}
Therefore, eqs.~(\ref{a4}) and (\ref{asymptotC}) 
show that the conformal blocks on the r.h.s.~of (\ref{c4}) 
should have the following asymptotic
behaviour when $b\to 0$:
\begin{equation}\boxed{
\label{defccb} {\cal F}\!\left(\Delta_4,\ldots,\Delta_1,\Delta,1+6Q^2\,|\,x\,\right) 
\; \sim \; \exp \left[ \frac{1}{b^2}\,f(\delta_4,\ldots,\delta_1,\delta \,|\,x\,) \right]}\;\;\;.
\end{equation}
The function 
$f(\delta_4,\ldots,\delta_1,\delta \,|\,x\,)$
is known as the classical 4-point block \cite{ZZ5}.
Taking in (\ref{c4}) the limit $ b\to 0$, one gets
\begin{equation}
\label{deltaaction} S^{\rm cl}(\delta_i,x;\delta)=S^{
\rm cl}(\delta_4,\delta_3,\delta) + S^{\rm cl}(\delta,\delta_2,\delta_1)
-f(x)-\bar f(\bar x).
\end{equation}
The classical limit of the left hand side of
(\ref{four:point:}) takes the form
$\textrm{e}^{-\frac{1}{b^2} S^{\rm cl}(\delta_4,\delta_3,\delta_2,\delta_1;x)}$, 
where
$$
S^{\rm cl}(\delta_4,\delta_3,\delta_2,\delta_1;x) \equiv
S^{\rm cl}(\delta_4,\delta_3,\delta_2,\delta_1;\infty,1,x,0).
$$
The classical limit of the right hand side of (\ref{four:point:}) 
is given by a saddle point approximation:
\begin{equation}
\label{siodlowa} \textrm{e}^{-\frac{1}{b^2} S^{\rm cl}(\delta_4,\delta_3,\delta_2,\delta_1;x)}=
\int\limits_0^\infty\!\textrm{d}p\; \textrm{e}^{-\frac{1}{b^2}S^{
\rm cl}(\delta_i,x;\delta)}\approx \textrm{e}^{-\frac{1}{b^2}S^{\rm cl}(\delta_i,x;\delta_s)},
\end{equation}
where 
$\delta_s ={\textstyle {1\over 4}} +p_s^2 $, and $p_s$
is the $s$-channel saddle point Liouville momentum, which is determined by the condition:
\begin{equation}
\label{saddle} {\partial \over \partial p}S^{\rm cl}\left(\delta_i,x;{\textstyle {1\over 4}}
+p^2\right)_{|p=p_s}=0.
\end{equation}
From (\ref{deltaaction}) and (\ref{siodlowa}) one gets thus the factorization
\begin{eqnarray}
\label{clasfact}
S^{\rm cl}(\delta_4,\delta_3,\delta_2,\delta_1;x) &=&
S^{\rm cl}(\delta_4,\delta_3,\delta_s(x)) + S^{\rm cl}(\delta_s(x),\delta_2,\delta_1)\nonumber \\
&&-f(\delta_4,\ldots,\delta_1,\delta_s(x)\,|\,x\,)
-\bar f(\delta_4,\ldots,\delta_1,\delta_s(x)\,|\,\bar x\,).
\end{eqnarray}
A few remarks are in order here. The asymptotical behaviour 
(\ref{defccb}) allows to calculate the classical 4-point block 
as a power expansion in $x$ order by order from the power expansion of the quantum Virasoro block.
Such expansion of  $f(\delta_4,\ldots,\delta_1,\delta \,|\,x\,)$ has been recently re-derived by Menotti 
\cite{Menotti:2014kra, Menotti:2016jut}, who has  employed a different method, i.e., based on the known 
connection between the classical 4-point block and the solution of the Riemann--Hilbert problem for the Heun equation.
The above-mentioned argumentation yielding the classical 
asymptotic of the 4-point block is an original idea of Zamolodchikovs 
spelled out in their  seminal paper \cite{ZZ5}. One can apply this idea to show an existence of the classical 
1-point block on the torus and calculate the power expansion of this function in the parameter 
$q=\exp 2\pi i\tau$, where $\tau$ is the torus modular parameter \cite{P1}.
The function $f(\delta_4,\ldots,\delta_1,\delta \,|\,x\,)$
represents the instanton contribution to the 
effective twisted superpotential of the supersymmetric ${\cal N} =2$, SU(2) gauge theory with four flavors; 
occurs in the computation of the 2-interval entanglement entropy in 2d CFT;
appears in the studies of various aspects and applications
of the AdS$_3$/CFT$_2 $ correspondence and quantum chaos.

As a final remark in this section let us stress that the exponential behaviour of spherical Virasoro blocks in the classical 
limit can be directly checked in multi-point cases as well. For instance, for the 5-point block 
(using Mathematica) one gets:\footnote{
Coefficients $f^{(2,0)}$ and $f^{(0,2)}$ are a bit more complicated. 
Note that variables $q_1$ and $q_2$
are not the positions of vertex operators in a correlation function (quantum block). 
The relationship between these variables is 
explained in subsection \ref{Mpc}.}
\begin{eqnarray*}
&&\lim_{b\to 0} b^2\log\mathcal{F}(\Delta_1,\ldots,\Delta_5,\Delta_{\text{in1}},\Delta _{\text{in2}},c\,|\,q_1,q_2)
\\
&&\hspace{3cm}=-\left(\delta _3+\delta _4+\delta _5-\delta _{\text{in1}}\right)\log q_1
-\left(\delta _4+\delta _5-\delta _{\text{in2}}\right)\log q_2
\\
&&\hspace{3cm}+f^{(1,0)}q_1+f^{(0,1)}q_2+f^{(1,1)}q_1q_2+f^{(2,0)}q_{1}^{2}+f^{(0,2)}q_{2}^{2}+\ldots,
\end{eqnarray*}
where
\begin{eqnarray*}
f^{(1,0)}(\delta_1,\ldots,\delta_5,\delta _{\text{in1}},\delta _{\text{in2}})
&=&-\frac{\left(\delta _1-\delta _2-\delta _{\text{in1}}\right) 
\left(\delta _3+\delta _{\text{in1}}-\delta _{\text{in2}}\right)}{2 \delta _{\text{in1}}},
\\
f^{(0,1)}(\delta_1,\ldots,\delta_5,\delta _{\text{in1}},\delta _{\text{in2}})&=&
\frac{\left(\delta _4-\delta _5+\delta _{\text{in2}}\right) \left(\delta _3-\delta _{\text{in1}}
+\delta _{\text{in2}}\right)}{2\delta _{\text{in2}}},
\\
f^{(1,1)}(\delta_1,\ldots,\delta_5,\delta _{\text{in1}},\delta _{\text{in2}})&=&
-\frac{\left(\delta _1-\delta _2-\delta _{\text{in1}}\right) \left(\delta _4-\delta _5+\delta _{\text{in2}}\right) \left(-\delta 
_3+\delta _{\text{in1}}+\delta _{\text{in2}}\right)}{4 \delta _{\text{in1}} \delta _{\text{in2}}},
\;\ldots\;.
\end{eqnarray*}

\section{Classical Virasoro blocks from Dotsenko--Fateev integrals}
\label{CVDF}
\subsection{Four--point case} 
Let us consider the Dotsenko--Fateev (DF) or 
Coulomb gas (CG) representation of some four-point 
conformal block on the sphere, namely,
\begin{equation}\label{ZDF}
{\cal Z}_{\rm DF}(\,\cdot\,|{\bf z}_f):=\left\langle 
V_{\hat\alpha_1}(0)V_{\hat\alpha_2}(x)
V_{\hat\alpha_3}(1)V_{\hat\alpha_4}(\infty)Q_{0,x}^{N_1}Q_{0,1}^{N_2}
\right\rangle,
\end{equation}
where ${\bf z}_f:=(0,x,1,\infty)$ and 
\begin{eqnarray}\label{V}
V_{\hat\alpha}(z)&:=& :{\rm e}^{\hat\alpha\phi(z)}:,\\
\label{Qc}
Q_{z_1,z_2}&:=&\int\limits_{z_1}^{z_2}V_{b}(u){\rm d}u
=\int\limits_{z_1}^{z_2}:\!{\rm e}^{b \phi(u)}\!:{\rm d}u.
\end{eqnarray}
The dot ``$\cdot$'' in 
${\cal Z}_{\rm DF}(\,\cdot\,|{\bf z}_f)$ stands for a dependence on $\hat\alpha_i$, $b$
and numbers of insertions $N_1$, $N_2$ of the screening operators (\ref{Qc}) defined
with different contours of integration.
Normal ordering of exponentials of a chiral free field within the matrix element (\ref{ZDF}) 
yields the following contour integral \cite{MMS}:
\begin{eqnarray}\label{ZDF2}
{\cal Z}_{\rm DF}(\,\cdot\,|{\bf z}_f)&=&
P(x)
\prod\limits_{\mu=1}^{N_1}\int\limits_{0}^{x}{\rm d}u_\mu
\prod\limits_{\mu=N_{1}+1}^{N_{1}+N_{2}}\int\limits_{0}^{1}{\rm d}u_\mu\nonumber
\\
&&\hspace{20pt}\times
\prod\limits_{\mu<\nu}\left(u_\nu - u_\mu\right)^{2\beta}
\prod\limits_{\mu}
u_{\mu}^{\alpha_1}\left(u_\mu-x\right)^{\alpha_2}\left(u_\mu-1\right)^{\alpha_3},
\end{eqnarray}
where 
$P(x):=x^\frac{\alpha_1\alpha_2}{2\beta}(1-x)^\frac{\alpha_2\alpha_3}{2\beta}$ and
$\beta:=b^2$, $\alpha_i:=2b\hat\alpha_i$.
It was not clear for a long time how to choose integration contours to get 
${\cal Z}_{\rm DF}(\,\cdot\,|{\bf z}_f)$ 
consistent with historically first Belavin--Polyakov--Zamolodchikov (BPZ) 
power series representation (\ref{block})-(\ref{blockBn}) of the four-point block.

Mironov, Morozov and Shakirov (MMS) showed in \cite{MMS} that the Dotsenko--Fateev integral 
(\ref{ZDF2}) precisely reproduces the BPZ four-point conformal block, i.e.,
\begin{equation}\label{MMS}\boxed{\;
{\cal Z}_{\rm DF}(\boldsymbol\alpha, N_1, N_2, \beta\,|\,{\bf z}_f)
\;=\;C_{\rm DF}\cdot x^{{\underline\Delta}-\Delta_1-\Delta_2}
\cdot {\sf B}\!\left(\boldsymbol\Delta,{\underline\Delta},c\,|\,x\right)\;}
\end{equation}
where $\boldsymbol\alpha:=\left\lbrace\alpha_1,\alpha_2,\alpha_3\right\rbrace$
and $\boldsymbol\Delta:=\left\lbrace\Delta_1,\Delta_2,\Delta_3,\Delta_4\right\rbrace$.
$C_{\rm DF}$ is the Dotsenko--Fateev normalization constant which does not depend
on $x$. 
It was found in \cite{MMS} that
\begin{equation}\label{CDF}
C_{\rm DF}=C_{N_1}(\alpha_1,\alpha_2)C_{N_2}(a,\alpha_3)
=C_{N_1}(2b\hat\alpha_1,2b\hat\alpha_2)C_{N_2}(a,2b\hat\alpha_3),
\end{equation}
where $a=\alpha_1+\alpha_2+2\beta N_1=2b(\hat\alpha_1+\hat\alpha_2)+2\beta N_1$ and
\begin{eqnarray}\label{CN}
C_{N}(p,t)&=&\prod\limits_{i=1}^{N}\int\limits_{0}^{1}{\rm d}u_i
\prod\limits_{i<j}(u_j-u_i)^{2\beta}\prod\limits_{i=1}^{N}u_{i}^{p}(u_i-1)^{t}\nonumber
\\
&=&
\prod\limits_{k=1}^{N}\frac{\Gamma(p+1+\beta(k-1))\Gamma(t+1+\beta(k-1))\Gamma(1+\beta k)}
{\Gamma(p+t+2+(N+k-2)\beta)\Gamma(\beta+1)}.
\end{eqnarray}
The block ${\sf B}\!\left(\boldsymbol\Delta,{\underline\Delta},c\,|\,x\right)$ in (\ref{MMS})
is given by\footnote{Here,
the $s$-channel four-point conformal block is
${\cal F}\!\left(\boldsymbol\Delta,{\underline\Delta},c\,|\,x\right)=x^{{\underline\Delta}-\Delta_1-\Delta_2}
{\sf B}\!\left(\boldsymbol\Delta,{\underline\Delta},c\,|\,x\right)$.
However, we will also refer to the function 
${\sf B}\!\left(\boldsymbol\Delta,{\underline\Delta},c\,|\,x\right)$ as the conformal block 
which should not lead to confusion. We follow here conventions used in \cite{MMS}
that differ from those in the previous section.
The block coefficients in (\ref{Bc}) can be written in our notation as 
$$
{\sf B}_{n}=\left\langle\,\beta^{\,n}_{ c,\underline{\Delta}_{s}}\!\left[\,^{\Delta_3}_{\Delta_4}\right]
\Big|\,\beta^{\,n}_{ c,\underline{\Delta}_{s}}\!\left[\,^{\Delta_1}_{\Delta_2}\right]\,\right\rangle
$$
(cf.~\cite{IMM}).}
\begin{eqnarray}\label{Bc}
{\sf B}\!\left(\boldsymbol\Delta,{\underline\Delta},c\,|\,x\right)&=&
1+\sum_{n=1}^\infty x^{n}
{\sf B}_{n}\!\left(\boldsymbol\Delta,{\underline\Delta},c\right)\nonumber
\\
&=&1+x\,\frac{(\underline\Delta+\Delta_2-\Delta_1)(\underline\Delta+\Delta_3-\Delta_4)}{2\underline\Delta}
+\ldots\;.
\end{eqnarray}
The identity (\ref{MMS}) holds provided that certain
relations between parameters are assumed, that is:
\begin{eqnarray*}
\Delta_i &=& \frac{\alpha_i\left(\alpha_i+2-2\beta\right)}{4\beta},\quad i=1,2,3,\\
\Delta_4 &=& \frac{\left(2\beta\left(N_1+N_2\right)+\alpha_1+\alpha_2+\alpha_3\right)
\left(2\beta\left(N_1+N_2\right)+\alpha_1+\alpha_2+\alpha_3+2-2\beta\right)}{4\beta},\\
{\underline\Delta} &=& \frac{\left(2\beta N_1+\alpha_1+\alpha_2\right)
\left(2\beta N_1+\alpha_1+\alpha_2+2-2\beta\right)}{4\beta},
\\
c &=& 1-6\left(\sqrt{\beta}-\frac{1}{\sqrt{\beta}}\right)^2,
\end{eqnarray*}
or in the alternative notation:
\begin{eqnarray}\label{rel}
\Delta_i &=& \hat\alpha_i\left(\hat\alpha_i+\frac{1}{b}-b\right),\quad i=1,2,3,
\nonumber\\
\Delta_4 &=& \left(b\left(N_1+N_2\right)+\hat\alpha_1+\hat\alpha_2+\hat\alpha_3\right)
\left(b\left(N_1+N_2\right)+\hat\alpha_1+\hat\alpha_2+\hat\alpha_3+\frac{1}{b}-b\right),
\nonumber\\
{\underline\Delta} &=& \left(bN_1+\hat\alpha_1+\hat\alpha_2\right)
\left(bN_1+\hat\alpha_1+\hat\alpha_2+\frac{1}{b}-b\right),
\nonumber\\
c &=& 1-6\left(b-\frac{1}{b}\right)^2.
\end{eqnarray}
Relations (\ref{rel}) were obtained by expanding ${\cal Z}_{\rm DF}(\,\cdot\,|{\bf z}_f)$ into power series,
\begin{eqnarray*}
{\cal Z}_{\rm DF}(\boldsymbol\alpha, N_1, N_2, \beta\,|\,{\bf z}_f) &=&
C_{\rm DF}\cdot x^{{\underline\Delta}-\Delta_1-\Delta_2}\cdot\left[1+\sum\limits_{n=1}^{\infty}x^n
{\cal J}_n\!\left(\boldsymbol\alpha, N_1, N_2, \beta\right)\right],
\end{eqnarray*}
and comparing coefficients ${\cal J}_n$ with ${\sf B}_n$. 
For more on the computation of ${\cal J}_k$, see~\cite{MMS} and \cite{MMS2}.

We aim to compute the (semi-classical) asymptotic of 
${\cal Z}_{\rm DF}(\,\cdot\,|{\bf z}_f)$ for $b\to\infty$. To do this, 
we introduce parameters $\eta$'s defined as follows:
\begin{equation}\label{etas}
\;\hat\alpha_i=b\eta_i\quad {\rm and} \quad\eta_i\sim{\cal O}(b^0), \quad i=1,2,3.\;
\end{equation}
The calculation of the asymptotical behaviour of 
${\cal Z}_{\rm DF}(\,\cdot\,|{\bf z}_f)$ for $b\to\infty$ can be made in two ways.

Indeed, note that on the one hand the asymptotic behaviour of 
${\cal Z}_{\rm DF}$ for $\hat\alpha_1$, $\hat\alpha_2$, $\hat\alpha_3$ 
in the form of (\ref{etas}) is dictated by the 
r.h.s. of the identity (\ref{MMS}) and is given by the classical limit of the BPZ four-point conformal block,
namely, 
\begin{eqnarray}\label{Zf}
\;{\cal Z}_{\rm DF}(\boldsymbol\alpha, N_1, N_2, b^2\,|\,{\bf z}_f) &\stackrel{b\to\infty}{\sim}&
\exp\Big[b^2\Big({\sf S}_{N_1}(2\eta_1,2\eta_2)+{\sf S}_{N_2}(2(\eta_1+\eta_2+N_1),2\eta_3)\nonumber
\\[3pt]
&&\hspace{1.cm}+\left(\underline{\delta}-\delta_1-\delta_2\right)\log x+
\hat f(\boldsymbol\delta,\underline{\delta}\,|\,x)\Big)\Big],
\end{eqnarray}
where 
\begin{equation}\label{delta} 
\underline{\delta}\;=\;\lim\limits_{b\to\infty}\frac{1}{b^2}\underline{\Delta}=
\left(N_1+\eta_1+\eta_2\right)\left(N_1+\eta_1+\eta_2-1\right)
\end{equation}
and 
$\boldsymbol\delta:=\left\lbrace\delta_1,\delta_2,\delta_3,\delta_4\right\rbrace$
with
\begin{eqnarray}\label{d1}
\delta_i &=& \lim\limits_{b\to\infty}\frac{1}{b^2}\Delta_i = \eta_i\left(\eta_i-1\right),
\quad i=1,2,3,\\
\label{d2}
\delta_4 &=& \lim\limits_{b\to\infty}\frac{1}{b^2}\Delta_4 = 
\left(N_1+N_2+\eta_1+\eta_2+\eta_3\right)\left(N_1+N_2+\eta_1+\eta_2+\eta_3-1\right)
\end{eqnarray}
as follows from (\ref{cl2}) and (\ref{rel}), (\ref{etas}).
The sum 
${\sf S}_{N_1}(2\eta_1,2\eta_2)+{\sf S}_{N_2}(2(\eta_1+\eta_2+N_1),2\eta_3)$ 
is nothing but the large $\beta$ ($\Leftrightarrow$ $b\to\infty$) 
asymptotic of $C_{\rm DF}$. Indeed, by making use of 
the method of steepest descent, or equivalently, using
known asymptotic formula for Gamma-function
from (\ref{CN}) one can derive (see appendix \ref{AppB})
\begin{equation}\label{LogCN}
\log C_{N}(b^2\lambda,b^2\omega)
\stackrel{b\to\infty}{\sim}b^{2}{\sf S}_{N}(\lambda,\omega).
\end{equation}
Then, from (\ref{CDF}) one can find 
\begin{eqnarray*}
\log C_{\rm DF}&=&\log C_{N_1}(2b\hat\alpha_1,2b\hat\alpha_2)+\log C_{N_2}(a,2b\hat\alpha_3)\\
&=&\log C_{N_1}\left(2b^2\eta_1,2b^2\eta_2\right)+\log C_{N_2}\left(2b^2(\eta_1+\eta_2+N_1),2b^2\eta_3\right)\\
& \stackrel{b\to\infty}{\sim} & 
b^2\Big({\sf S}_{N_1}(2\eta_1,2\eta_2)+{\sf S}_{N_2}(2(\eta_1+\eta_2+N_1),2\eta_3)\Big).
\end{eqnarray*}
The sum of the third and fourth terms in the exponent on the r.h.s. of (\ref{Zf}) is nothing but the 
classical four-point block introduced in (\ref{defccb}), i.e.,
\begin{equation}\label{fcl}
f(\boldsymbol\delta,\underline{\delta}\,|\,x)\;=\;
\left(\underline{\delta}-\delta_1-\delta_2\right)\log x+
\hat f(\boldsymbol\delta,\underline{\delta}\,|\,x),
\end{equation}
where the function 
$\hat f(\boldsymbol\delta,\underline{\delta}\,|\,x)$ 
in conventions employed here reads as follows
\begin{eqnarray}\label{fclS}
\hat f(\boldsymbol\delta,\underline{\delta}\,|\,x)
&=&
\lim\limits_{b\to\infty}
\frac{1}{b^2}\log\left(1+\sum_{n=1}^\infty x^{n}
{\sf B}_{n}\!\left(\boldsymbol\Delta,{\underline\Delta},c\right)\right)\nonumber\\[5pt]
&=&
x\,\frac{(\underline\delta+\delta_2-\delta_1)(\underline\delta+\delta_3-\delta_4)}{2\underline\delta}
+\ldots\;.
\end{eqnarray}
Let us note that the calculation of the asymptotic (\ref{Zf}) is similar to a derivation of 
the classical limit of the DOZZ Liouville 4-point function (see subsection \ref{Classical limit}). 
Also here, i.e. in (\ref{Zf}), we have a decomposition onto the sum of functions ${\sf S}_{N}(\cdot,\cdot)$
and the classical block. The function ${\sf S}_{N}(\cdot,\cdot)$ can be seen as an analogue 
of the classical 3-point Liouville action (\ref{cl}).

On the other hand, given (\ref{etas}), one can write 
\begin{eqnarray*}
{\cal Z}_{\rm DF}(\,\cdot\,|{\bf z}_f) &=&
x^{2b^2\eta_1\eta_2}(1-x)^{2b^2\eta_2\eta_3}
\prod\limits_{\mu=1}^{N_1}\int\limits_{0}^{x}{\rm d}u_\mu
\prod\limits_{\mu=N_{1}+1}^{N_{1}+N_{2}}\int\limits_{0}^{1}{\rm d}u_\mu\nonumber\\
&&\hspace{100pt}\times
\prod\limits_{\mu<\nu}\left(u_\nu-u_\mu\right)^{2b^2}\prod\limits_{\mu}
u_{\mu}^{2b^2\eta_1}
\left(u_\mu -x\right)^{2b^2\eta_2}
\left(u_\mu-1\right)^{2b^2\eta_3}
\\
&=&
{\rm e}^{b^2\left(2\eta_1\eta_2\log x+2\eta_2\eta_3\log(1-x)\right)}
\prod\limits_{\mu=1}^{N_1}\int\limits_{0}^{x}{\rm d}u_\mu
\prod\limits_{\mu=N_{1}+1}^{N_{1}+N_{2}}\int\limits_{0}^{1}{\rm d}u_\mu\nonumber\\
&&\hspace{-60pt}\times
\exp\left\lbrace -b^2\left[-2\sum\limits_{\mu<\nu}\log(u_\nu-u_\mu)
-\sum\limits_{\mu=1}^{N_1+N_2}\left[2\eta_1\log u_{\mu}
+2\eta_2\log\left(u_{\mu}-x\right)
+2\eta_3\log\left(u_{\mu}-1\right)\right]\right]
\right\rbrace\\
&=:&
{\rm e}^{b^2\left(2\eta_1\eta_2\log x+2\eta_2\eta_3\log(1-x)\right)}
\;I_{\rm DF}(\,\cdot\,|{\bf z}_f),
\end{eqnarray*}
where
$$
I_{\rm DF}(\,\cdot\,|{\bf z}_f)\;:=\;\prod\limits_{\mu=1}^{N_1}\int\limits_{0}^{x}{\rm d}u_\mu
\prod\limits_{\mu=N_{1}+1}^{N_{1}+N_{2}}\int\limits_{0}^{1}{\rm d}u_\mu\,
{\rm e}^{-b^2\,W(\,\cdot\,|{\bf z}_f,{\bf u})},\quad\quad
{\bf u}:=\left\lbrace u_1,\ldots,u_{N_1+N_2=N}\right\rbrace,
$$
and 
\begin{eqnarray}\label{Wp}
W(\,\cdot\,|{\bf z}_f,{\bf u})
&:=& -2\sum\limits_{\mu<\nu}\log(u_\nu-u_\mu)\nonumber
\\
&&-\sum\limits_{\mu=1}^{N_1+N_2}\left[2\eta_1\log u_{\mu}
+2\eta_2\log\left(u_{\mu}-x\right)
+2\eta_3\log\left(u_{\mu}-1\right)\right].
\end{eqnarray}
In the limit $b\to\infty$
the integral $I_{\rm DF}(\,\cdot\,|{\bf z}_f)$ is given by the saddle point value.
The saddle point equations take the form:
\begin{eqnarray}\label{sadd}
\frac{\partial W(\,\cdot\,|{\bf z}_f,{\bf u})}{\partial u_{\mu}}\;=\;0\,
\;\;\Leftrightarrow\;\;
\frac{2\eta_1}{u_\mu}+\frac{2\eta_2}{u_\mu-x}+\frac{2\eta_3}{u_\mu-1}
+\sum\limits_{\nu\neq\mu}^{N}\frac{2}{u_\mu-u_\nu}\;=\;0,&&
\\
\forall\;\mu=1,\ldots, N=N_1+N_2.&&\nonumber
\end{eqnarray}
So, in the limit $b\to\infty$ one obtaines
\begin{equation}\label{Zsad}
\;Z_{\rm DF}(\,\cdot\,|{\bf z}_f)\;\sim\;
\frac{\left(\frac{2\pi}{b^2}\right)^{\frac{N}{2}}}{\sqrt{\det{\bf W}_{N}}}
\exp\Big[-b^2\Big(W(\,\cdot\,|{\bf z}_f,{\bf u}^{\rm c})
-2\eta_1\eta_2\log x-2\eta_2\eta_3\log(1-x)\Big)\Big]\;.
\end{equation}
Here, ${\bf u}^{\rm c}:=\left\lbrace u_{1}^{\rm c},\ldots,u_{N}^{\rm c}\right\rbrace$ is a solution of 
eqs.~(\ref{sadd}) and
$$
{\bf W}_{N}\;:=\;
\left(\frac{\partial^2 W}{\partial u_\mu\partial u_\nu}\right)\Big|_{{\bf u}={\bf u}^{\rm c}},
\quad\quad
\mu,\nu=1,2,\ldots,N
$$
is the Hessian matrix of the exponent $W(\,\cdot\,|{\bf z}_f,{\bf u})$
(see appendix \ref{AppA}). 

Combining (\ref{Zf}) with (\ref{Zsad}) and taking $b\to\infty$ one comes 
into the following conjecture.

\medskip\noindent
{
The classical four-point block on the sphere $f(\boldsymbol\delta,\underline{\delta}\,|\,x)$, i.e., the
function specified in (\ref{fcl})-(\ref{fclS}), where  $\underline{\delta}$ is the
intermediate classical weight of the form (\ref{delta}) and 
$\delta_i$, $i=1,\ldots,4$ are the external classical conformal 
weights given by (\ref{d1})-(\ref{d2}), is determined by the critical value 
$W(\,\cdot\,|{\bf z}_f,{\bf u}^{\rm c})$ of the ``action'' (\ref{Wp}). 
Precisely,\footnote{Asymptotics (\ref{Zf}) and (\ref{Zsad}) yield (\ref{fW}) plus
$\left[b^{-2}\log\left((2\pi/b^2)^{N/2}\right)
-b^{-2}\log\left(\sqrt{\det{\bf W}_{N}}\right)\right]\to 0 \;\;{\rm for}\;\;b\to\infty.$
Interestingly, gauge theory analogues of classical blocks, i.e., instanton twisted superpotentials are calculated 
in the Nekrasov--Shatashvili limit also as critical values of some functions obtained from the Nekrasov instanton partition 
functions in the contour integral representations.}
\begin{eqnarray}\label{fW}
f(\boldsymbol\delta,\underline{\delta}\,|\,x)&=&
-\,W(\,\cdot\,|{\bf z}_f,{\bf u}^{\rm c})
-\Big({\sf S}_{N_1}(2\eta_1,2\eta_2)+{\sf S}_{N_2}(2(\eta_1+\eta_2+N_1),2\eta_3)\Big)\nonumber
\\[4pt]
&&+\,2\eta_1\eta_2\log x+2\eta_2\eta_3\log(1-x).
\end{eqnarray}}

\subsection{Multi--point case}
\label{Mpc}
A multi-point counterpart of the relation (\ref{MMS}) takes the form 
\cite{MMS}
\begin{equation}\label{MMSmpoint}\boxed{\,
{\cal Z}_{\rm DF}(\boldsymbol\alpha, {\bf N}, \beta\,|\,{\bf q})
\;=\;C_{\rm DF}\cdot \prod_{i=1}^{m-3} q_{i}^{{\rm deg}_{i}}
\cdot {\sf B}\!\left(\boldsymbol\Delta,
\underline{\boldsymbol\Delta},c\,|\,{\bf q}\right)\,}
\end{equation}
where 
$$
\boldsymbol\alpha:=\left\lbrace\alpha_1,\alpha_2,\ldots,\alpha_{m-1}\right\rbrace,
\quad\quad
{\bf N}:=\left\lbrace N_1,\ldots,N_{m-2}\right\rbrace,
$$
$$
\boldsymbol\Delta:=\left\lbrace\Delta_1,\ldots,\Delta_{m}\right\rbrace,
\quad\quad
\underline{\boldsymbol\Delta}:=\left\lbrace\underline\Delta_1,\ldots,\underline\Delta_{m-3}\right\rbrace,
\quad\quad
{\bf q}:=\left\lbrace q_1,\ldots,q_{m-3}\right\rbrace.
$$
Let us stress that in (\ref{MMSmpoint}) new coordinates 
$q_1,\ldots,q_{m-3}$ are used in comparison with (\ref{MMS}).
These variables are related to the locations 
$0,x_1,\ldots,x_{m-3},1,\infty$ of the vertex operators in the chiral correlator by the formulae:
\begin{equation}\label{xvq}
x_1=q_1q_2\ldots q_{m-3},
\quad
x_2=q_2q_3\ldots q_{m-3},
\quad
x_i=\prod_{j=i}^{m-3}q_j,
\quad
\ldots,
\quad
x_{m-3}=q_{m-3}.
\end{equation}
In addition, it is assumed that $x_0=q_0=0$ and $x_{m-2}=q_{m-2}=1$,
in accordance with the SL$(2,\mathbb{C})$ symmetry of the theory. 
A reason for using the new coordinates $q_1,\ldots, q_{m-3}$ is that in these variables 
the multi-point blocks have more convenient expansions in positive powers. 

Explicitly, the l.h.s. of eq.~(\ref{MMSmpoint}) looks as follows \cite{MMS}
\begin{eqnarray}\label{ZDFm}
{\cal Z}_{\rm DF}(\boldsymbol\alpha, {\bf N}, \beta\,|\,{\bf q})&:=&\left\langle
\prod\limits_{r=0}^{m-2}V_{\hat\alpha_{r+1}}(x_r)
\prod\limits_{r=1}^{m-2}Q_{0,x_r}^{N_r}
\right\rangle\nonumber
\\
&=&
\left\langle
\prod\limits_{r=0}^{m-2}:{\rm e}^{\hat\alpha_{r+1}\phi(x_r)}:
\prod\limits_{r=1}^{m-2}\left(\int\limits_{0}^{x_r}:\!{\rm e}^{b \phi(u)}\!:{\rm d}u\right)^{N_r}
\right\rangle\nonumber\\
&=&
\prod\limits_{\mu<\nu}\left(x_\nu-x_\mu\right)^{\frac{\alpha_\mu \alpha_{\nu}}{2\beta}}
\prod\limits_{r=1}^{m-2}\prod\limits_{\mu=1}^{N_r}\int\limits_{0}^{x_r}{\rm d}u_{N_1+\ldots+N_{r-1+\mu}}\nonumber
\\
&&\hspace{20pt}\times\,
\prod\limits_{\mu<\nu}\left(u_\nu-u_\mu\right)^{2\beta}
\prod\limits_{\mu=1}^{N_1+\ldots+N_{m-2}}\prod\limits_{r=0}^{m-2}
\left(u_\mu-x_r\right)^{\alpha_{r+1}},
\end{eqnarray}
where as before $\beta:=b^2$, $\alpha_i:=2b\hat\alpha_i$. 
For $m=4$ the formula (\ref{ZDFm}) yields (\ref{ZDF2}). For $m=5$ we have
\begin{eqnarray*}
{\cal Z}_{\rm DF}&=&q_{1}^{\frac{\alpha_1(\alpha_2+\alpha_3)}{2\beta}}
q_{2}^{\frac{\alpha_1\alpha_2}{2\beta}}
\left(1-q_1\right)^{\frac{\alpha_3\alpha_4}{2\beta}}
\left(1-q_1q_2\right)^{\frac{\alpha_2\alpha_4}{2\beta}}
\left(1-q_2\right)^{\frac{\alpha_2\alpha_3}{2\beta}}\\
&&\times\prod\limits_{\mu=1}^{N_1}\int\limits_{0}^{q_1q_2}{\rm d}u_\mu
\prod\limits_{\mu=1}^{N_2}\int\limits_{0}^{q_2}{\rm d}u_{N_1+\mu}
\prod\limits_{\mu=1}^{N_3}\int\limits_{0}^{1}{\rm d}u_{N_1+N_2+\mu}\\
&&\times\prod\limits_{\mu<\nu}\left(u_\nu-u_\mu\right)^{2\beta}
\prod\limits_{\mu=1}^{N_1+N_2+N_3}u_{\mu}^{\alpha_1}\left(u_{\mu}-q_1q_2\right)^{\alpha_2}
\left(u_{\mu}-q_1\right)^{\alpha_3}\left(u_{\mu}-1\right)^{\alpha_4}.
\end{eqnarray*}
On the r.h.s. of eq.~(\ref{MMSmpoint}) the constant $C_{\rm DF}$ is now given by the product
\begin{equation}\label{CDFmul}
C_{\rm DF}=C_{N_1}(\alpha_1,\alpha_2)C_{N_2}(\underline{\Delta}_{1},\alpha_3)
\ldots C_{N_{m-2}}(\underline{\Delta}_{m-3},\alpha_{m-1}).
\end{equation}
The exponent ${\rm deg}_{i}$ is of the form
${\rm deg}_{i}=\underline{\Delta}_{i}-\Delta_{1}-\ldots-{\Delta}_{i+1}$.
The $m$-point block ${\sf B}\!\left(\boldsymbol\Delta,\underline{\boldsymbol\Delta},c\,|\,{\bf q}\right)$ 
in (\ref{MMSmpoint}) is calculated in the ``comb'' channel (see Fig.3).
\begin{figure}[tbp]
\label{comb}
\centering
\includegraphics[height=2.3cm, width=10.cm]{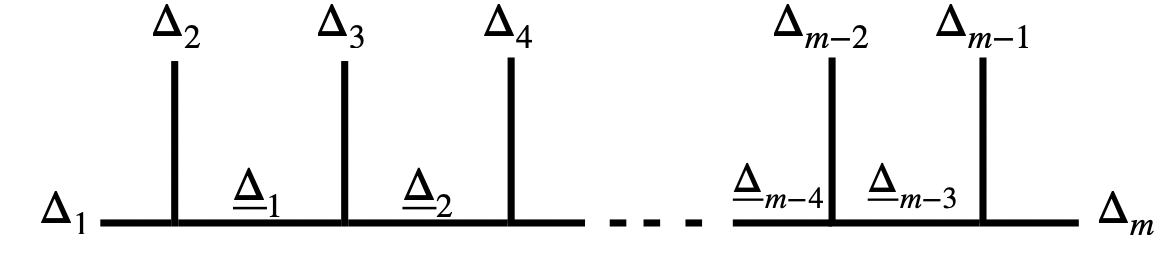}
\caption{\it On the above diagram chiral vertex operators with external conformal weights 
$\boldsymbol\Delta:=\left\lbrace\Delta_1,\ldots,\Delta_{m}\right\rbrace$ have 
the following locations, $V_{\Delta_1}$ is at $x_0=0$, $V_{\Delta_{m-2}}$ is at 
$x_{m-3}$, $V_{\Delta_{m-1}}$ is at $x_{m-2}=1$ and $V_{\Delta_{m}}$ is at $x_{m-1}=\infty$.}
\end{figure}
Again, the identity (\ref{MMSmpoint}) is fulfilled when certain 
relationships between parameters take place, namely,
\begin{eqnarray*}
\Delta_i &=& \frac{1}{4\beta}\alpha_i\left(\alpha_i+2-2\beta\right),\quad i=1,\ldots,m-1,
\\[6pt]
\Delta_m &=& \frac{1}{4\beta}
\left(\alpha_1+\ldots+\alpha_{m-1}+2\beta\left(N_1+\ldots+N_{m-2}\right)\right)
\\
&&\times\;
\left(\alpha_1+\ldots+\alpha_{m-1}+2\beta\left(N_1+\ldots+N_{m-2}\right)+2-2\beta\right),
\\[6pt]
{\underline\Delta}_i &=& \frac{1}{4\beta}
\left(\alpha_1+\ldots+\alpha_{i+1}+2\beta\left(N_1+\ldots+N_{i}\right)\right)
\\
&&\times\;
\left(\alpha_1+\ldots+\alpha_{i+1}+2\beta\left(N_1+\ldots+N_{i}\right)+2-2\beta\right),
\\[6pt]
c &=& 1-6\left(\sqrt{\beta}-\frac{1}{\sqrt{\beta}}\right)^2
\end{eqnarray*}
or
\begin{eqnarray}\label{relmpoint}
\Delta_i &=& \hat\alpha_i\left(\hat\alpha_i+\frac{1}{b}-b\right),\quad i=1,\ldots,m-1,
\nonumber\\[6pt]
\Delta_m &=& \left(b\left(N_1+\ldots+N_{m-2}\right)+\hat\alpha_1+\ldots+\hat\alpha_{m-1}\right)\nonumber
\\
&&\times\;
\left(b\left(N_1+\ldots+N_{m-2}\right)+\hat\alpha_1+\ldots+\hat\alpha_{m-1}+\frac{1}{b}-b\right),
\nonumber\\[6pt]
{\underline\Delta}_p &=& \left(b\left(N_1+\ldots+N_{p}\right)+\hat\alpha_1+\ldots+\hat\alpha_{p+1}\right)\nonumber
\\
&&\times\;
\left(b\left(N_1+\ldots+N_{p}\right)+\hat\alpha_1+\ldots+\hat\alpha_{p+1}+\frac{1}{b}-b\right),
\nonumber\\
&&p=1,\ldots,m-3,\nonumber
\\[6pt]
c &=& 1-6\left(b-\frac{1}{b}\right)^2.
\end{eqnarray}

As in the four-point case there are two ways to calculate the asymptotic behaviour of 
${\cal Z}_{\rm DF}(\boldsymbol\alpha, {\bf N},b^2\,|\,{\bf q})$ with
$\boldsymbol\alpha:=\left\lbrace 2b^2\eta_1,\ldots, 2b^2\eta_{m-1}\right\rbrace$
for $b\to\infty$. Repeating the same calculation steps as in the previous subsection 
one gets in the multi-point case an extension of the conjectured relation (\ref{fW}). 
We formulate this result below.
{
\begin{enumerate}
\item
Let us introduce
\begin{itemize}
\item[(i)]
the ``action''
\begin{eqnarray}\label{Wmulti}
W\left(\,\cdot\,|\,{\bf x}({\bf q}),{\bf u}\right)
&:=& -2\sum\limits_{\mu<\nu}\log(u_\nu-u_\mu)\nonumber
\\
&-&\sum\limits_{\mu=1}^{N_1+\ldots+N_{m-2}}\left[\sum\limits_{r=0}^{m-2} 
2\eta_{r+1}\log\left(u_{\mu}-x_{r}({\bf q})\right)\right],
\end{eqnarray}
where ${\bf x}:=\left\lbrace x_{0}=0, x_{1},\ldots, x_{m-3}, x_{m-2}=1\right\rbrace$,
${\bf u}:=\left\lbrace u_{1},\ldots,u_{N=N_1+\ldots+N_{m-2}}\right\rbrace$ and the symbol
$x_{r}({\bf q})$ means that each $x_r$, $r=0,\ldots,m-2$ is replaced with
corresponding $q$'s, in accordance with (\ref{xvq}) and $x_0=q_0=0$, $x_{m-2}=q_{m-2}=1$;
\item[(ii)]
the critical value
$W\left(\,\cdot\,|\,{\bf x}({\bf q}),{\bf u}^{\rm c}\right)$ of (\ref{Wmulti}),
i.e., the function (\ref{Wmulti}) evaluated on the stationary solution 
${\bf u}^{\rm c}:=\left\lbrace u_{1}^{\rm c},\ldots,u_{N=N_1+\ldots+N_{m-2}}^{\rm c}\right\rbrace$ 
of the saddle point equations:
\begin{eqnarray}\label{saddM}
\frac{\partial W\left(\,\cdot\,|\,{\bf x}({\bf q}),{\bf u}\right)}{\partial u_{\mu}}\;=\;0\,
\;\;\Leftrightarrow\;\;
\sum\limits_{r=0}^{m-2}\frac{2\eta_{r+1}}{u_\mu-x_{r}({\bf q})}
+\sum\limits_{\nu\neq\mu}^{N}\frac{2}{u_\mu-u_\nu}\;=\;0,&&
\\
\forall\;\mu=1,\ldots, N=N_1+\ldots+N_{m-2}.&&\nonumber
\end{eqnarray}
\end{itemize}
\item
Let ${\sf S}_{N}(\cdot,\cdot)$ denotes an exponent in the large $\beta$ asymptotic (\ref{LogCN})
of the structure constant (\ref{CN}) given by (\ref{SNcl}). 
\item 
Let ${\hat f}(\boldsymbol\delta,\underline{\boldsymbol\delta}\,|\,{\bf q})$ 
be a series part of the $m$-point classical Virasoro block in the ``comb'' channel, i.e.,
\begin{equation}\label{mfclS}
{\hat f}(\boldsymbol\delta,\underline{\boldsymbol\delta}\,|\,{\bf q})\;=\;\lim\limits_{b\to\infty}
\frac{1}{b^2}\log{\sf B}\!\left(\boldsymbol\Delta,\underline{\boldsymbol\Delta},c\,|\,{\bf q}\right),
\end{equation}
where
\begin{itemize}
\item[(a)]
the parameters $\boldsymbol\Delta$, $\underline{\boldsymbol\Delta}$, $c$ of the quantum block
${\sf B}\!\left(\boldsymbol\Delta,\underline{\boldsymbol\Delta},c\,|\,{\bf q}\right)$ are given by (\ref{relmpoint}); 
\item[(b)]
the parameters of the classical block are
$\boldsymbol\delta:=\left\lbrace\delta_1,\ldots,\delta_m\right\rbrace$,
$\underline{\boldsymbol\delta}:=\left\lbrace\underline{\delta}_1,\ldots,
\underline{\delta}_{m-3}\right\rbrace$,
\begin{eqnarray}\label{d1M}
\delta_i &=& \lim\limits_{b\to\infty}\frac{1}{b^2}\Delta_i = \eta_i\left(\eta_i-1\right),
\quad i=1,\ldots,m-1,\\
\label{d2M}
\delta_m &=& \lim\limits_{b\to\infty}\frac{1}{b^2}\Delta_m = 
\left(N_1+\ldots+N_{m-2}+\eta_1+\ldots+\eta_{m-1}\right)\nonumber
\\
&&\hspace{1.8cm}\times
\left(N_1+\ldots+N_{m-2}+\eta_1+\ldots+\eta_{m-1}-1\right),\\
\underline{\delta}_p &=& \lim\limits_{b\to\infty}\frac{1}{b^2}\underline{\Delta}_p =
\left(N_1+\ldots+N_p+\eta_1+\ldots+\eta_{p+1}\right)\nonumber\\
&&\hspace{1.8cm}\times\left(N_1+\ldots+N_p+\eta_1+\ldots+\eta_{p+1}-1\right),\nonumber
\\
&&\quad p=1,\ldots,m-3.
\end{eqnarray}
\end{itemize}
\item
We conjecture that 
\begin{eqnarray}\label{ConM}
{\hat f}(\boldsymbol\delta,\underline{\boldsymbol\delta}\,|\,{\bf q})&=&
-W\left(\,\cdot\,|\,{\bf x}({\bf q}),{\bf u}^{\rm c}\right)+
\sum\limits_{\mu<\nu}2\eta_{\mu}\eta_{\nu}\log(x_{\nu}({\bf q})-x_{\mu}({\bf q}))\nonumber
\\
&&-\Big({\sf S}_{N_1}(2\eta_1,2\eta_2)+
{\sf S}_{N_2}(\underline{\delta}_{1},2\eta_3)+\ldots+
{\sf S}_{N_{m-2}}(\underline{\delta}_{m-3},2\eta_{m-1})\Big)\nonumber
\\[4pt]
&&-\sum\limits_{i=1}^{m-3}(\underline{\delta}_{i}-\delta_{1}-\ldots-{\delta}_{i+1})\log q_i.
\end{eqnarray}
\end{enumerate}}

\noindent
The statement is that the part 
${\hat f}(\boldsymbol\delta,\underline{\boldsymbol\delta}\,|\,{\bf q})$ 
of the $m$-point classical Virasoro block given by the series in variables $q_1,\ldots,q_{m-3}$, which
is calculable order by order from the quantum block expansion via (\ref{mfclS}), 
can be summed up to the r.h.s. of  eq.~(\ref{ConM}). On the r.h.s. the contributions are: 
($a$) the ``on-shell action'' $W\left(\,\cdot\,|\,{\bf x}({\bf q}),{\bf u}^{\rm c}\right)$;
($b$) the exponent ${\sf S}_{N_1}+{\sf S}_{N_2}+\ldots+{\sf S}_{N_{m-2}}$ 
in the ``classical asymptotic'' of $C_{\rm DF}$, where this time $C_{\rm DF}$ is given by (\ref{CDFmul});
($c$) the classical limit of the prefactor $q_{1}^{{\rm deg}_{1}}\ldots q_{m-3}^{{\rm deg}_{m-3}}$.

\section{Richardson and Gaudin models}
\label{RGmodels}
\subsection{Richardson's solution}
The Hamiltonian (\ref{RBCSH}) was studied by 
Richardson long ago with intent to be used in nuclear physics. 
$\hat{\rm H}_{\rm rBCS}$ 
can be written in terms of the so-called ``hard-core'' boson operators 
$b_{j}^{\dagger}=c_{j+}^{\dagger}c_{j-}^{\dagger}$ and $b_{j}=c_{j-}c_{j+}$
which create, annihilate fermion pairs, respectively and obey the following 
commutation rules
$\big[b_{j},b_{j'}^{\dagger}\big]=\delta_{j,j'}(1-2\hat{\rm N}_{j})$,
$\hat{\rm N}_{j}=b_{j}^{\dagger}\,b_{j}$.
The reduced BCS Hamiltonian
rewritten in terms of these operators reads as follows
\begin{equation}\label{HrBCS}
\hat{\rm H}_{\rm rBCS}=\sum\limits_{j}2\varepsilon_{j}b_{j}^{\dagger}b_{j}
-gd\sum\limits_{j,j'}b_{j}^{\dagger}b_{j'}.
\end{equation}
Sums in (\ref{HrBCS}) run over a set of doubly degenerate energy levels with 
energies $\varepsilon_j$, $j=1,\ldots,\Omega$.
In the 1960s
Richardson exactly solved an eigenvalue problem for (\ref{HrBCS})
through the Bethe ansatz \cite{R1,RS,RS2}. (For a pedagogical introduction to the 
Richardson solution, see \cite{vonDelft:1999si}.) Specifically,
Richardson proposed an ansatz for an exact eigenstate of the Hamiltonian 
(\ref{HrBCS}), namely,
$$
\ket{N}=\prod\limits_{\nu=1}^{N}B^{\dagger}_{\nu}\ket{0}\;,
$$
where the pair operators $B^{\dagger}_{\nu}$ have 
the form appropriate to the solution of the one-pair problem, i.e.,
$$
B^{\dagger}_{\nu}=\sum\limits_{j=1}^{\Omega}\frac{b^{\dagger}_{j}}{2\varepsilon_j-u_{\nu}}\;.
$$
Here every energy level is either empty or occupied by a pair of fermions.
The quantities $u_{\nu}$ are pair energies. They are understood as auxiliary parameters which 
are then chosen to fulfill the eigenvalue equation
\begin{equation}\label{Hsp}
\hat{\rm H}_{\rm rBCS}\,\ket{N}={\rm E}_{\rm rBCS}(N)\,\ket{N}\;,
\end{equation}
where ${\rm E}_{\rm rBCS}(N)\!=\!\sum_{\nu=1}^{N}u_{\nu}$. 
Indeed, the state  $\ket{N}$ is an eigenstate of 
the pairing Hamiltonian (\ref{HrBCS}) if the $N$ pair energies $u_{\nu}$ 
are, complex in general, solutions of the (Bethe ansatz) equations:
\begin{eqnarray}\label{Req}
\frac{1}{gd}+\sum\limits_{i=1}^{\Omega}\frac{1}{u_\nu-z_i}
=\sum\limits_{\mu\neq\nu}^{N}\frac{2}{u_\nu-u_\mu}\;\;\;\;
{\rm for}\;\;\nu=1,\ldots,N,
\end{eqnarray}
where $z_i=2\varepsilon_{i}$.
(For a derivation of eqs.~(\ref{Req}) 
using a commutator technique, see \cite{vonDelft:1999si}.)
The ground state of the system is given by the solution of eqs.~(\ref{Req}) which
gives the lowest value of ${\rm E}_{\rm rBCS}(N)$. 
The normalized state $\ket{N}$ can be written in the following form
\begin{equation}\label{Rs}
\ket{N}=\frac{C}{\sqrt{N!}}\sum\limits_{j_1,\ldots,j_N}\psi(j_1,\ldots,j_N)
b^{\dagger}_{j_1}\ldots b^{\dagger}_{j_N}\ket{0}.
\end{equation}
In (\ref{Rs}) the normalization constant $C$ is determined by the condition $\bracket{N}{N}\!=\!1$ while
$\psi(j_1,\ldots,j_N)$ is the Richardson wave function of the form
\begin{equation}\label{RsP}
\psi^{\rm R}(j_1,\ldots,j_N)=
\sum\limits_{\mathcal{P}}\prod\limits_{k=1}^{N}\frac{1}{z_{j_k}-u_{{\cal P}_k}}.
\end{equation}
The sum in (\ref{RsP}) runs over all permutations $\mathcal{P}$ of $1,\ldots,N$.

There is a connection between the Richardson (reduced BCS) model and a class of integrable
spin models obtained by Gaudin \cite{Gaudin1,Gaudin2}.
In 1976 Gaudin proposed the so-called rational, trigonometric and elliptic integrable 
models based on sets of certain commuting Hamiltonians.
The simplest model in this family it is the rational model 
which is defined by a collection of Hamiltonians of the form
\begin{eqnarray}\label{Gaudin}
\hat{\mathrm{H}}_{{\rm G}, i} &=& \sum\limits_{j\neq i}^{\Omega}
\frac{1}{\varepsilon_i-\varepsilon_j}
\left[{\rm t}_{i}^{0}{\rm t}_{j}^{0}
+\frac{1}{2}\left({\rm t}_{i}^{+}{\rm t}_{j}^{-}+{\rm t}_{i}^{-}{\rm t}_{j}^{+}\right)\right]
\nonumber
\\
&=:&
\sum\limits_{j\neq i}^{\Omega}
\frac{{\bf t}_i\cdot{\bf t}_j}{\varepsilon_i-\varepsilon_j}.
\end{eqnarray}
$\hat{\mathrm{H}}_{{\rm G}, i}$ describe a system of 
interacting spins labeled by $i=1,\ldots,\Omega$.
Each separate spin corresponds to a spin-$\frac{1}{2}$
realization of the $\mathfrak{su}(2)$ algebra generated by 
${\rm t}^{0}$, ${\rm t}^{+}$, ${\rm t}^{-}$.\footnote{
The spin-$\frac{1}{2}$ generators 
$\left\lbrace {\rm t}^{0}, {\rm t}^{+}, {\rm t}^{-}\right\rbrace$ of the $\mathfrak{su}(2)$ algebra
are built out of the standard ones $\left\lbrace {\rm t}^{x}, {\rm t}^{y},{\rm t}^{z}\right\rbrace$
which obey $\left[{\rm t}^a,{\rm t}^b\right]=i\epsilon_{abc}{\rm t}^c$, i.e., 
${\rm t}^0={\rm t}^z$, ${\rm t}^+={\rm t}^x+i{\rm t}^y$, ${\rm t}^-={\rm t}^x-i{\rm t}^y$. 
This yields:
$$\left[{\rm t}^{0}, {\rm t}^{+}\right]\;=\;{\rm t}^+, 
\quad\quad
\left[{\rm t}^{0}, {\rm t}^{-}\right]\;=\;-{\rm t}^-,
\quad\quad
\left[{\rm t}^{+}, {\rm t}^{-}\right]\;=\;2{\rm t}^0.
$$ 
For a system of $\Omega$ spins
this leads to a set of generators satisfying
$\left[{\rm t}_{j}^{0}, {\rm t}_{k}^{\pm}\right]=\pm\delta_{jk}{\rm t}_{k}^{\pm}$
and
$\left[{\rm t}_{j}^{+}, {\rm t}_{k}^{-}\right]=2\delta_{jk}{\rm t}_{k}^{0}$.
The quadratic Casimir operator 
which commutes with the generators is given by 
${\bf t}\cdot{\bf t}={\rm t}^{0}{\rm t}^{0}
+\frac{1}{2}\left({\rm t}^{+}{\rm t}^{-}+{\rm t}^{-}{\rm t}^{+}\right).$}
Interestingly, the spin-$\frac{1}{2}$ generators can be written in terms of
the hard-core boson operators, i.e., ${\rm t}_{j}^{+}=b_{j}^{+}$, ${\rm t}_{j}^{-}=b_{j}$,
${\rm t}_{j}^{0}=\frac{1}{2}-\hat{\rm N}_j$.
Therefore, the Gaudin Hamiltonians $\hat{\mathrm{H}}_{{\rm G}, i}$
can be diagonalized by means of the Richardson method. As before the energy eigenvalue is given by
${\rm E}_{{\rm G},i}(N)=\sum_{\nu=1}^{N}u_{\nu}$, but this time the parameters $u_{\nu}$
satisfy
\begin{eqnarray}\label{Geq}
\sum\limits_{j=1}^{\Omega}\frac{1}{u_{\nu}-z_j}
&=&
\sum\limits_{\mu\neq\nu}^{N}\frac{2}{u_{\nu}-u_{\mu}}
\;\;\;\;
{\rm for}\;\;\nu=1,\ldots,N.
\end{eqnarray}
Eqs.~(\ref{Geq}) are nothing 
but the Richardson eqs.~(\ref{Req}) in the limit $g\To\infty$.

In 1997 Cambiagio, Rivas and Saraceno (CRS) \cite{Cambiaggio:1997vz} 
found that conserved charges of the Richardson--BCS model,
i.e., mutually commuting operators $\hat{\rm R}_i$ which commute with the 
reduced BCS Hamiltonian (\ref{HrBCS}), 
are given in terms of the rational Gaudin Hamiltonians:
\begin{eqnarray}\label{Rii}
\hat{\rm R}_i&=&-{\rm t}_{i}^{0}-gd\,\hat{\mathrm{H}}_{{\rm G}, i}\nonumber
\\
&=&
-{\rm t}_{i}^{0}-gd\sum\limits_{j\neq i}^{\Omega}
\frac{{\bf t}_i\cdot{\bf t}_j}{\varepsilon_{i}-\varepsilon_{j}},
\;\;\;\;\;\;\;\; i=1,\ldots,\Omega.
\end{eqnarray}
The quantum integrals 
of motion (\ref{Rii}) itself can be seen as a set of commuting Hamiltonians. These are 
famous Gaudin magnets known also as the central spin model (cf.~\cite{Claeys:2018zwo} and refs therein).
The Gaudin magnets play a crucial role in the study of integrable spin models
and their connections with 2d CFT, see for instance \cite{Babu,BabuFlume}. 
Moreover, there exist several examples of the integrable (time-dependent) 
Hamiltonians, relevant for concrete physical applications, 
which can be reduced to the Gaudin magnets, and  integrability of which is 
determined by the celebrated Knizhnik--Zamolodchikov equation, cf.~eg.~\cite{Yuzbashyan:2018gbu}.

The fact that operators $\hat{\rm R}_i$ commute with $\hat{\rm H}_{\rm rBCS}$ 
implies that $\hat{\rm H}_{\rm rBCS}$ can be expressed in terms of $\hat{\rm R}_i$, i.e.,
\begin{equation}\label{HH}
\hat{\rm H}_{\rm rBCS}\;=\;\hat{\rm H}_{\rm XY} 
+ \sum\limits_{j=1}^{\Omega}\varepsilon_{j}+gd\left(\tfrac{1}{2}\Omega-N\right),
\end{equation}
where
\begin{equation}\label{HH2}
\hat{\rm H}_{\rm XY}\;=\;\sum\limits_{j=1}^{\Omega}2\varepsilon_{j}\hat{\rm R}_{j}
+gd\Big(\sum\limits_{j=1}^{\Omega}\hat{\rm R}_j\Big)^2
-\frac{3}{4}gd\,\Omega,
\end{equation}
cf.~\cite{Sierra:1999mp,Cambiaggio:1997vz}. 
The intermediate step that leads to 
(\ref{HH})-(\ref{HH2}) is that the operator $\hat{\rm H}_{\rm XY}$ 
can be expressed as a spin chain Hamiltonian,
\begin{equation}\label{Hxy}
\hat{\rm H}_{\rm XY}\;=\;-\sum\limits_{j=1}^{\Omega}2\varepsilon_{j}{\rm t}^{0}_{j}
-\frac{gd}{2}\left({\rm T}^{+}{\rm T}^{-}+
{\rm T}^{-}{\rm T}^{+}\right).
\end{equation}
The matrices ${\rm T}^{a}=\sum_{j=1}^{\Omega}{\rm t}_{j}^{a}$, 
$a\in\lbrace 0,+,-\rbrace$ satisfy the ${\mathfrak{su}(2)}$ 
algebra $\left[{\rm T}^{a},{\rm T}^{b}\right]=f_{c}^{ab}\,{\rm T}^{b}$, where
\begin{equation}\label{fsu2}
f_{\,+}^{\,+\,0}=f_{-}^{\,0\,-}=-1\,, 
\quad
f_{0}^{\,+\,-}=2\,.
\end{equation}
The Casimir of that algebra is given by
${\bf T}\cdot {\bf T}={\rm T}^{0}{\rm T}^{0}
+\frac{1}{2}({\rm T}^{+}{\rm T}^{-}+{\rm T}^{-}{\rm T}^{+})$. 
Eqs.~(\ref{HH}) and (\ref{HH2}) open a possibility to calculate eigenvalues of quantum integrals 
of motion $\hat{\rm R}_i$ by solving them with respect to $\hat{\rm R}_i$. 
Unluckily, CRS did not compute the eigenvalues of $\hat{\rm R}_i$. Perhaps, they 
didn't know about the Richardson solution.

In fact, for a long time Richardson results 
were not assimilated by the scientific community until the end of the 20th century, 
when it was realized that the Richardson model describes pairing of time-reversed 
electrons in metalic mesoscopic grains.
On this new wave of interest in the Richardson solution 
more novel ideas sailed out. In particular, in those days Sierra proposed
in the aforementioned paper \cite{Sierra:1999mp} 
a link between 2d CFT and the exact solution, and integrability of the reduced BCS model.
As was explained, for example, in \cite{Sierra:2001cx} there were several motivations that led 
to the formulation of such links. One of them was the so-called electrostatic analogy mentioned 
in the introduction. 

Indeed, the Richardson equations (\ref{Req}) can be interpreted within a two-dimensional
electrostatic picture. Let us consider a set of $\Omega$ charges with charge $q=-1$ fixed 
at the positions $z_i=2\epsilon_i$ on the real axis and an uniform field parallel to this axis
with strength $-\frac{1}{gd}$. The task is to find equilibrium positions of $N$ charges with 
charge $q=+2$ at positions $u_\nu$ subject to their mutual repulsion and an attraction wit $\Omega$
charges with $q=-1$, and an action of the uniform field. 
(The constant electric field is absent in the Gaudin model.) 
The electrostatic potential is given by $U+\bar U$,  where the holomorphic part reads as 
follows
\begin{eqnarray}\label{U}
U({\bf z},{\bf u}) 
&=&
-\sum\limits_{i<j}^{\Omega}\log(z_i-z_j)
-4\sum\limits_{\nu<\mu}^{N}\log(u_\nu-u_\mu)\nonumber\\
&+&
2\sum\limits_{i=1}^{\Omega}\sum\limits_{\nu=1}^{N}\log(z_i-u_\mu)
+\frac{1}{gd}\Big(-\sum\limits_{i=1}^{\Omega}z_i+2\sum\limits_{\nu=1}^{N}u_{\nu}\Big)\,.
\end{eqnarray}
Here
${\bf z}:=\lbrace z_1,\ldots,z_\Omega\rbrace$ and 
${\bf u}:=\lbrace u_1,\ldots,u_N\rbrace$.
Therefore, eqs.~(\ref{Req}) are nothing but the equilibrium conditions:
\begin{equation}
\frac{\partial U({\bf z},{\bf u})}{\partial u_{\nu}}=0\,,
\quad\quad \nu=1,\ldots,N\;.
\end{equation} 

As already mentioned, the electrostatic picture can be immersed in conformal field theory. 
The relevant CFT here is given by the $\widehat{\mathfrak{su}(2)}_{k}$ 
WZW model in the so-called critical level limit $k\To -2$.
This embedding led to a new result, namely, Sierra using CFT methods found closed 
expression for eigenvaluse of conserved charges (\ref{Rii}), i.e.,
\begin{eqnarray}\label{Li}
\lambda_i &=& \frac{gd}{2}
\frac{\partial U({\bf z},{\bf u}^{\sf c})}{\partial z_i}\Big|_{z_i=2\varepsilon_i}\nonumber
\\
&=&
-\frac{1}{2}+gd\left(\sum\limits_{\nu=1}^{N}\frac{1}{2\varepsilon_{i}-u_{\nu}^{\sf c}}
-\frac{1}{4}\sum\limits_{j\neq i}^{\Omega}\frac{1}{\varepsilon_{i}-\varepsilon_{j}}\right).
\end{eqnarray} 
The quantity $U({\bf z},{\bf u}^{\sf c})\equiv U_{\sf c}$ in eq.~(\ref{Li}) above is the 
critical value of the potential (\ref{U}), i.e., $U({\bf z},{\bf u})$ evaluated
on the solution ${\bf u}^{\sf c}:=\left\lbrace u_{1}^{\sf c},\ldots,u_{N}^{\sf c}\right\rbrace$
of the Richardson equations (\ref{Req}). 
We review a derivation of eq.~(\ref{Li}) in subsection \ref{KZeqInt}.
Before that we will briefly recall the conformal field theory construction presented in
\cite{Sierra:1999mp} (see also \cite{Sierra:2001cx}).

\subsection{Embedding in conformal field theory}
To realize the Richardson solution within 2d CFT 
and derive expression for eigenvalues of CRS conserved charges
Sierra proposed to consider a {\it perturbed} affine conformal block 
\begin{eqnarray}\label{fcg}
\psi^{\rm CG}_{\bf m}({\bf z}):=
\left\langle{\sf V}_{gd}
\Phi_{m_1}^{\frac{1}{2}}(z_1)\ldots\Phi_{m_\Omega}^{\frac{1}{2}}(z_{\Omega})
\tilde\Phi_{j_{\Omega+1}}^{j_{\Omega+1}}(\infty)
\oint\limits_{C_1}{\rm d}u_{1} {\sf S}(u_1)\ldots
\oint\limits_{C_N}{\rm d}u_{N} {\sf S}(u_N)
\right\rangle
\end{eqnarray}
which contains 
\begin{itemize} 
\item[---] the $\widehat{\mathfrak{su}(2)}_{k}$ WZW chiral primary fields in a free field realization:\footnote{
The building blocks of the free field realization are:
(a) the $\beta\gamma$ system, i.e.,
the system of two bosonic fields which satisfy the following OPEs: 
$$
\gamma(z)\beta(w)\sim\frac{1}{(z-w)}\,,
\;\;\;\;\;\;\;
\beta(z)\gamma(w)\sim -\frac{1}{(z-w)}
$$
and have conformal weights: $\Delta_\beta=1$, $\Delta_\gamma=0$;
(b) Virasoro primary chiral vertex operators represented as 
normal ordered exponentials of the chiral free field $\phi(z)$ having conformal weights (\ref{Dwzw}).
For details on the $\widehat{\mathfrak{su}(2)}_{k}$ WZW model and its operator representation, see \cite{DiFMS}.}
$$
\Phi_{m}^{j}(z)=\left(\gamma(z)\right)^{j-m}V_{\alpha}(z)\,,
\quad
\alpha=(k+2)^{-\frac{1}{2}}j=-2\alpha_0 j\,
$$
\begin{equation}\label{Dwzw}
\Delta_{\alpha}=\alpha(\alpha-2\alpha_0)=\frac{j(j+1)}{k+2};
\end{equation}
\item[---] WZW screening charges:
$$
{\sf Q}=\oint\limits_{C}\frac{{\rm d}u}{2\pi i}\,{\sf S}(u)\,,
\quad\quad
{\sf S}(u)=\beta(u)V_{2\alpha_0}(u)\,;
$$
\item[---] the operator
\vspace{-1cm}$$
{\sf V}_{gd}=\exp\left(-\frac{i\alpha_0}{gd}\oint\limits_{C_g}{\rm d}zz\partial_z\phi(z)\right),
$$
which breaks conformal invariance.
\end{itemize}
Within this realization to every energy level $z_i=2\varepsilon_i$ 
is associated the field $\Phi_{m_i}^{j}(z_i)$ 
with the ``spin'' $j=\frac{1}{2}$ and the ``third component'' $m_i=\frac{1}{2}$
(or $m_i=-\frac{1}{2}$) if the corresponding energy level is empty (or occupied) by a pair of fermions.
Integration variables $u_{\nu}$ in screening operators correspond to   
Richardson parameters. The operator ${\sf V}_{gd}$ 
implements the coupling $gd$ and is a source of the term $\frac{1}{gd}$ in 
the Richardson equations. In the next subsection, following \cite{Sierra:1999mp}, we spell out how this idea works, i.e., 
how this construction is related to the Richardson solution and how it solves eigenproblems for (\ref{Rii}).

\subsection{Knizhnik--Zamolodchikov equation and integrability}
\label{KZeqInt}
In CFT framework the integrability of the Richardson model is determined by the
Knizhnik--Zamolodchikov (KZ) equation,
\begin{equation}\label{KZEsu2}
\left(\kappa\partial_{z_i}-\sum\limits_{j\neq i}^{\Omega+1}
\frac{{\bf t}_i\cdot{\bf t}_j}{z_i-z_j}\right)\Psi^{\rm WZW}(z_1,\ldots,z_{\Omega+1})\;=\;0\,,
\end{equation}
which is obeyed by the $\widehat{\mathfrak{su}(2)}_{k}$ WZW conformal block
$\Psi^{\rm WZW}(z_1,\ldots,z_{\Omega+1})$.
The coupling constant $\kappa$ in eq.~(\ref{KZEsu2}) is 
\begin{equation}\label{kappa}
\kappa\;=\;\frac{(k+2)}{2}\,,
\end{equation}
where $k$ is the level of the Kac--Moody algebra of the WZW model.\footnote{See \cite{DiFMS}} 
Here, ${\bf t}_i$ are the $\mathfrak{su}(2)$ matrices in the $j_i$ 
representation acting at the $i$-th site.\footnote{For a derivation of the KZ equation in general case, 
see \cite{DiFMS}.} 
Let us observe that for $z_{\Omega+1}\To\infty$ the term 
$({\bf t}_i\cdot{\bf t}_{\Omega+1})/(z_i-z_{\Omega+1})$ in (\ref{KZEsu2}) equals zero.
Therefore, in such a case one can use the identity\footnote{Here we take the $\Omega+1$--point conformal block 
with $z_{\Omega+1}=\infty$ to make contact with calculations performed in \cite{Sierra:1999mp}.}
$$
\frac{1}{2gd}\hat{\rm R}_i + \frac{1}{2gd}{\rm t}_{i}^{0}=-\sum\limits_{j\neq i}^{\Omega}
\frac{{\bf t}_i\cdot{\bf t}_j}{z_{i}-z_{j}},
$$ 
which follows from the form of $\hat{\rm R}_i$ for $z_j=2\varepsilon_j$, and the KZ 
equation (\ref{KZEsu2}) rewrite to the following form 
\begin{equation}\label{KZEsu2II}
\left(\kappa\partial_{z_i}+\frac{1}{2gd}\hat{\rm R}_i + \frac{1}{2gd}{\rm t}_{i}^{0}
\right)\Psi^{\rm WZW}(z_1,\ldots,z_{\Omega},\infty)=0.
\end{equation}
Let us define a new function $\Psi$ related to $\Psi^{\rm WZW}$ by 
\begin{equation}\label{def}
\Psi^{\rm WZW}=\exp\left(\frac{1}{2gd\kappa}\hat{\rm H}_{\rm XY}\right)\Psi.
\end{equation}
Elementary calculations yield 
\begin{eqnarray*}
\kappa\partial_{z_i}\Psi^{\rm WZW}&=&\exp\left(\frac{1}{2gd\kappa}\hat{\rm H}_{\rm XY}\right)\nonumber
\\
&\times& 
\left[\frac{1}{2gd}\left(\partial_{z_i}\hat{\rm H}_{\rm XY}\right)\Psi+\kappa\partial_{z_i}\Psi\right].
\end{eqnarray*}
The operators $\hat{\rm R}_i$ and $\hat{\rm H}_{\rm XY}$ commute therefore on can write  
$$
\hat{\rm R}_i\exp\left(\frac{1}{2gd\kappa}\hat{\rm H}_{\rm XY}\right)
=\exp\left(\frac{1}{2gd\kappa}\hat{\rm H}_{\rm XY}\right)\hat{\rm R}_i.
$$
Using these two identities one gets from eq.~$(\ref{KZEsu2II})$ the equation for $\Psi$,
$$
\left(\kappa\partial_{z_i}+\frac{1}{2gd}\partial_{z_i}\hat{\rm H}_{\rm XY}
+\frac{1}{2gd}\hat{\rm R}_i + \frac{1}{2gd}{\rm t}_{i}^{0}
\right)\Psi=0.
$$
Let us observe that $\partial_{z_i}\hat{\rm H}_{\rm XY}=-{\rm t}_{i}^{0}$ as it follows from (\ref{Hxy}) 
for $z_j=2\varepsilon_j$. Therefore, we have
\begin{equation} \label{KZPsi}
\frac{1}{2gd}\hat{\rm R}_{i}\Psi=-\kappa\partial_{z_i}\Psi.
\end{equation}
Eq.~(\ref{KZPsi}) is completely equivalent to the KZ 
equation and has been obtained in \cite{Sierra:1999mp}. 
Our aim now is to calculate its classical limit 
$c\To\infty\;\iff\;\alpha_0\To\infty$.
We will see that in the limit $\alpha_0\To\infty$ eq.~(\ref{KZPsi}) is nothing but the eigenvalue problem
for conserved charges $\hat{\rm R}_{i}$.
This result we summarize in the following lemma.

If the function $\Psi({\bf z})$, ${\bf z}:=\lbrace z_1,\ldots,z_{\Omega}\rbrace$ 
in (\ref{KZPsi}) has the following asymptotical behaviour 
in the limit $\alpha_{0}\To\infty$,
\begin{equation}\label{Psisem}
\Psi({\bf z})\sim\tilde\psi({\bf z})
\,{\rm e}^{-\alpha_{0}^{2}U({\bf z},{\bf u}^{\sf c})},
\end{equation}
where $\tilde\psi({\bf z})$ has the ``lightness'' property
\begin{equation}\label{light}
\lim\limits_{\kappa\to 0}\kappa\partial_{z_i}\tilde\psi({\bf z})\;=\;0
\end{equation}
then the function $\tilde\psi({\bf z})$ solves the eigenvalue problem for 
the Richardson model conserved charges,
\begin{equation}\label{eigenR}
\hat{\rm R}_{i}\tilde\psi\;=\;\lambda_{i}\tilde\psi
\end{equation}
with the eigenvalues $\lambda_i$ of integrals of motion given by 
\begin{equation}\label{lam}
\lambda_i\;=\;\frac{gd}{2}\frac{\partial U_{\sf c}}{\partial z_i}\;.
\end{equation}

Indeed, a substitution of (\ref{Psisem}) into eq.~(\ref{KZPsi}) gives 
\begin{equation}\label{KZcl}
\frac{1}{2gd}\hat{\rm R}_{i}\tilde\psi+
\kappa\partial_{z_i}\tilde\psi-
\alpha_{0}^{2}\kappa(\partial_{z_i}U_{\sf c})\tilde\psi\;=\;0.
\end{equation}
The central charge $c$ in the case under consideration is parametrized in terms of
the background charge\footnote{What is background charge $\alpha_{0}$, 
see \cite{DiFMS}.} $\alpha_{0}$ and is related to $k$ --- 
the level of the Kac--Moody algebra:
$$
c=3-12\alpha_{0}^{2}=\frac{3k}{k+2}.
$$
In addition, $k$ is related to the coupling $\kappa$ in the KZ equation via (\ref{kappa}).
So the large central charge limit $c\To\infty$ corresponds to $\alpha_{0}\To\infty$ and to the critical 
level limit $k\To -2$, and finally to $\kappa\To 0$. Knowing that one can find that
\begin{eqnarray}\label{ka}
\alpha_{0}^{2}\kappa &=& \frac{\alpha_{0}^{2}}{2}2\kappa=\frac{\alpha_{0}^{2}}{2}(k+2)
=\frac{\alpha_{0}^{2}}{2}\frac{3k}{3-12\alpha_{0}^{2}}\nonumber
\\
&=&\frac{3k}{\frac{6}{\alpha_{0}^{2}}-24}
\To\frac{-6}{-24}=\frac{1}{4}
\end{eqnarray}
for  $\alpha_{0}\To\infty$ and simultaneously $k\To -2$. Therefore, in the limit 
$\alpha_{0}\To\infty$ taking into account (\ref{light}) from (\ref{KZcl}) we obtain
$\hat{\rm R}_{i}\tilde\psi=\frac{1}{2}gd(\partial_{z_i}U_{\sf c})\tilde\psi$.

Thus, to solve spectral problems for CRS conserved charges one has to
construct the function $\Psi({\bf z})$ 
related to WZW conformal block via (\ref{def}) and having the $\alpha_{0}\To\infty$
asymptotic of the form (\ref{Psisem}).
This concrete realization of $\Psi({\bf z})$ with required asymptotic  
behaviour was found in \cite{Sierra:1999mp}. Precisely,
$\Psi({\bf z})$ is nothing but the perturbed WZW block (\ref{fcg}).
The latter has the asymptotic (\ref{Psisem}) for $\alpha_0\To\infty$.\footnote{
It therefore seems that the asymptotics of the type (\ref{cla2}) are more general 
and appear not only for Virasoro blocks, but also, as here, for perturbed and unperturbed affine WZW blocks.}
Indeed, from (\ref{fcg}) one gets
$$
\psi^{\rm CG}_{\bf m}({\bf z})\;=\;\oint\limits_{C_1}{\rm d}u_{1}\ldots
\oint\limits_{C_N}{\rm d}u_{N}\,
\psi^{\beta\gamma}_{\bf m}({\bf z},{\bf u}){\rm e}^{-\alpha_{0}^{2}\,U({\bf z},{\bf u})}\;,
$$
where
$$
\psi^{\beta\gamma}_{\bf m}({\bf z},{\bf u})\;=\;\left\langle
\prod\limits_{i=1}^{\Omega}\gamma^{\frac{1}{2}-m_i}(z_i)
\prod\limits_{\nu=1}^{N}\beta(u_\nu)\beta^{s+2j_{\Omega+1}}(\infty)\gamma^{2j_{\Omega+1}}(\infty)
\right\rangle\;
$$
and $U({\bf z},{\bf u})$ is the potential (\ref{U}).
In the limit $\alpha_0\To\infty$ the above multiple contour integral
can be calculated using the saddle point method. Recall, the stationary solutions of $U$ 
are given by the solutions of the Richardson equation. As a result one gets (\ref{Psisem}), where
$\tilde\psi({\bf z})=\psi^{\beta\gamma}_{\bf m}({\bf z},{\bf u}^{\sf c})$ is the Richardson wave function
and $U_{\sf c}\equiv U({\bf z},{\bf u}^{\sf c})$ is the ``on-shell'' potential appearing in (\ref{Li}). $U_{\sf c}$ is
named the Coulomb energy in \cite{Sierra:1999mp}.


One sees that the Coulomb energy $U_{\sf c}$ and 
eigenvalues (\ref{Li}) of CRS conserved charges depend on the Richardson parameters 
${\bf u}^{\sf c}:=\left\lbrace u_{1}^{\sf c},\ldots,u_{N}^{\sf c}\right\rbrace$.
It would be nice to have techniques that allow to find $U_{\sf c}$ without need to solve 
the Richardson equations. It seems possible to develop such methods. 
Let us note that $U_{\sf c}$ is an analogue of the ``on-shell'' action 
$W(\,\cdot\,|\,\cdot\,,{\bf u}^{\rm c})$ appearing in relations (\ref{fW}) and (\ref{ConM}). 
In the unperturbed case $g\To\infty$ corresponding to the rational Gaudin model $U_{\sf c}$ is a direct analogue. Indeed, for 
$g\To\infty$  the VEV (\ref{fcg}) is nothing but the WZW block, 
where an integrand decomposes onto two standard CFT contributions, 
i.e., some Virasoro block and a chiral function of the $\beta\gamma$ system.
So, it is reasonable to expect that $U_{\sf c}$ should be related to the classical limit of some WZW
block in unperturbed case and to its ``perturbed counterpart'' for finite coupling $g$.

\section{Concluding remarks}\label{Con}
In the present work, exploiting known connections between power 
series and integral representations of Virasoro conformal blocks 
(the MMS identities \cite{MMS}), 
we have proposed new expressions (\ref{fW}) and (\ref{ConM}) for 
classical Virasoro blocks.
The four-point identity (\ref{fW}) we analyzed numerically in some 
special case (see appendix \ref{AppC}).
We observed that for some real solution of the saddle point 
equations, the 
real parts of both sides of this relation are almost identical for $x$ 
in the vicinity of zero. Unfortunately, the imaginary parts of this 
relationship differ, which may be due to the incompatibility of 
branches of functions appearing on both sides. A more comprehensive 
analytical and numerical analysis of these relationships we leave as a 
topic for further research.

When deriving formulae (\ref{fW}) and (\ref{ConM}), we have been inspired by 
the ideas used in the work \cite{Sierra:1999mp}. 
Also this work was our special motivation for this research line. 
As has been mentioned in the last paragraph of the previous section, 
identities connecting the classical limit of conformal blocks expansions with 
the saddle point approximation of Coulomb gas integrals can lead to new 
computational techniques applicable to the solution of the Richardson and Gaudin models. 
It seems to be a highly non trivial task to develop such methods for the Richardson model. 
Here in fact one needs to derive the MMS-type relations for the perturbed WZW block (\ref{fcg}).\footnote{
This can be difficult because due to a broken conformal symmetry standard 2d CFT tools 
such as factorization on intermediate states perhaps cannot be used.
There is also a problem of choosing appropriate integration contours, 
which are crucial in the MMS relations, and which would not spoil the results obtained in \cite{Sierra:1999mp}.}
However, this should be achievable for the Gaudin model, 
where a relevant tool is the (unperturbed) WZW block in a free field representation. 
Interestingly, it seems that in both cases matrix models tools can help. In particular, 
the Dijkgraaf--Vafa quasiclassical approach (see e.g.~\cite{MSh}) 
and techniques developed in \cite{EyMar} by Eynard and Marchal should be helpful.

Regardless of the possible applications, the use of matrix models 
technology in the study of the classical limit of Virasoro and WZW conformal blocks appears promising.
This is the topic that we also intend to re-examine soon.

\appendix
\section{Saddle point asymptotics of complex valued integrals}
\label{AppA}
In this appendix we attached a key tool for our calculations, i.e., a 
steepest descent formula for the $N$-dimensional complex integrals proved by N.~Bleistein in 
the note:~``Saddle Point Contribution for an $n$-fold Complex-Valued Integral''.\footnote{Available on
[{\rm http://citeseerx.ist.psu.edu/viewdoc/download?doi=10.1.1.661.8737\&rep=rep1\&type=pdf}].}

Let
\begin{equation}
{\cal K}_{N}(\omega)\;=\;\int{\rm d}z_1\ldots{\rm d}z_{N}f({\bf z})\,{\rm e}^{-\omega\Psi({\bf z})},
\quad\quad
{\bf z}=\left\lbrace z_1,\ldots,z_N\right\rbrace,
\end{equation}
where it is assumed that
\begin{list}{}{\itemindent=2mm \parsep=0mm \itemsep=0mm \topsep=0mm}
\item[(i)] the exponent has a {\it simple} saddle point in all variables that is
\begin{equation}\label{seq}
\boldsymbol\nabla_z\Psi({\bf z})\;=\;\boldsymbol 0,
\quad\quad {\bf z}={\bf z}_{\rm sad}:=\left\lbrace z_{1,\rm sad},\ldots,z_{N,\rm sad}\right\rbrace;
\end{equation}
\item[(ii)] the Hessian matrix $\boldsymbol\Psi_N$, the matrix of second derivatives of $\Psi({\bf z})$,
is non-singular at the simple saddle point determined by (\ref{seq}), i.e.,
\begin{equation}
\det\left(\boldsymbol\Psi_N\right)\neq 0,
\quad\quad
\boldsymbol\Psi_N=\left(\frac{\partial^2\Psi}{\partial z_i\partial z_j}\right)\Big|_{{\bf z}={\bf z}_{\rm sad}},
\quad\quad
i,j=1,2,\ldots,N.
\end{equation}
Then, the leading order asymptotic of ${\cal K}_{N}(\omega)$ for $\omega\to\infty$ is
\begin{equation}
{\cal K}_{N}(\omega)\;\sim\;\left(\frac{2\pi}{\omega}\right)^{\frac{N}{2}}
\frac{f({\bf z}_{\rm sad})}{\sqrt{\det\left(\boldsymbol\Psi_N\right)}}\,
{\rm e}^{-\omega\Psi({\bf z}_{\rm sad})}.
\end{equation}
\end{list}

\section{Asymptotics of Selberg integral}
\label{AppB}
The formula:
\begin{eqnarray}\label{Selberg}
I_{N}(\lambda_1,\lambda_2,\lambda_3)&:=&\int\limits_{0}^{1}{\rm d}t_{1}\ldots\int\limits_{0}^{1}{\rm d}t_{N}
\prod\limits_{j=1}^{N}t_{j}^{\lambda_1}\left(1-t_j\right)^{\lambda_2}
\prod\limits_{1\leq j<i\leq N}\left(t_i-t_j\right)^{2\lambda_3}\nonumber
\\
&=&\prod\limits_{k=0}^{N-1}
\frac{\Gamma(\lambda_1+1+k\lambda_3)\Gamma(\lambda_2+1+k\lambda_3)\Gamma(1+(k+1)\lambda_3)}
{\Gamma(\lambda_1+\lambda_2+2+(N+k-1)\lambda_3)\Gamma(1+\lambda_3)}\nonumber
\\
&=&\prod\limits_{\ell=1}^{N}
\frac{\Gamma(\lambda_1+1+(\ell-1)\lambda_3)\Gamma(\lambda_2+1+(\ell-1)\lambda_3)\Gamma(1+\ell\lambda_3)}
{\Gamma(\lambda_1+\lambda_2+2+(N+\ell-2)\lambda_3)\Gamma(1+\lambda_3)},
\end{eqnarray}
where $0<t_1<\ldots<t_N<1$, is known as the Selberg integral (see 
\cite{Forr,Var}).
In the case $N=1$ this integral is the Euler beta integral:
$$
\int\limits_{0}^{1}t^{\lambda_1}(1-t)^{\lambda_2}{\rm d}t
\;=\;\frac{\Gamma(\lambda_1+1)\Gamma(\lambda_2+1)}
{\Gamma(\lambda_1+\lambda_2+2)}.
$$

Let
$$
\lambda_1=\frac{\sf a}{\kappa},
\quad
\lambda_2=\frac{\sf b}{\kappa},
\quad
\lambda_3=\frac{\sf c}{\kappa},
\quad
{\sf a},{\sf b},{\sf c},\kappa>0
\quad{\rm and}\quad\kappa\to 0.
$$
For the left hand side of (\ref{Selberg}) the method of steepest descent gives (see appendix \ref{AppA})
\begin{equation}\label{sadd1}
I_{N}\left(\lambda_1,\lambda_2,\lambda_3\right)\;\;\sim\;\;(2\pi\kappa)^\frac{N}{2}\;
\Phi\left(t^0;\frac{\sf a}{\kappa},\frac{\sf b}{\kappa},\frac{\sf c}{\kappa}\right)\;
\left[{\rm Hess}(-S\left(t^0;{\sf a},{\sf b},{\sf c}\right))\right]^{-\frac{1}{2}}.
\end{equation}
Here, 
$$
\Phi(t;\lambda_1,\lambda_2,\lambda_3)=\prod\limits_{j=1}^{N}t_{j}^{\lambda_1}\left(1-t_j\right)^{\lambda_2}
\prod\limits_{1\leq j<i\leq N}\left(t_i-t_j\right)^{2\lambda_3}
$$
and
\begin{eqnarray*}
S\left(t;{\sf a},{\sf b},{\sf c}\right)&=&
\kappa\log\Phi\left(t;\frac{\sf a}{\kappa},\frac{\sf b}{\kappa},\frac{\sf c}{\kappa}\right)
\\
\br &=&\sum\limits_{j=1}^{N}\left({\sf a}\log t_j + {\sf b}\log(1-t_j)\right)
+\sum\limits_{1\leq j<i\leq N}2{\sf c}\log\left(t_i-t_j\right);
\end{eqnarray*}
$t^0$ is the critical point of $S$ in $\left\lbrace t\in \mathbb{R}^N \;|\;0<t_1<\ldots<t_N<1\right\rbrace$ and
\begin{eqnarray*}
{\rm Hess}(-S)&=&\det\left(-\frac{\partial^2 S}{\partial t_i\partial t_j}\right),
\\
-\frac{\partial^2 S}{\partial t_{i}^{2}}&=&{\sf a}\frac{1}{t_{i}^{2}}+{\sf b}\frac{1}{(t_{i}-1)^2}+2{\sf c}\sum\limits_{j\neq i}
\frac{1}{(t_{i}-t_{j})^2},
\\
-\frac{\partial^2 S}{\partial t_{i} \partial t_{j}}&=&-2{\sf c}\frac{1}{(t_{i}-t_{j})^2}.
\end{eqnarray*}
Equations for critical points are
$$
\frac{\partial S}{\partial t_{i}}=0\quad\Leftrightarrow\quad
\frac{\sf a}{t_i}+\frac{\sf b}{t_i-1}+\sum\limits_{j\neq i}\frac{2 \sf c}{t_i-t_j}=0,
\quad
i=1,\ldots,N.
$$
Let 
$$
\lambda_{N}=t^{0}_{1}\ldots t^{0}_{N},
\quad\quad
\mu_N=(1-t^{0}_{1})\ldots (1-t^{0}_{N}),
\quad\quad
\delta=\prod\limits_{i<j}\left(t^{0}_{i}-t^{0}_{j}\right)^2
$$
then
$$
\Phi\left(t^0;\frac{\sf a}{\kappa},\frac{\sf b}{\kappa},\frac{\sf c}{\kappa}\right)
=\lambda_{N}^{\frac{\sf a}{\kappa}}\cdot\mu_{N}^{\frac{\sf b}{\kappa}}
\cdot\delta^{\frac{\sf c}{\kappa}}
\quad\Longrightarrow\quad
S\left(t^{0};{\sf a},{\sf b},{\sf c}\right)={\sf a}\log\lambda_N+{\sf b}\log\mu_N+{\sf c}\log\delta.
$$
So, the asymptotic (\ref{sadd1}) takes the form
\begin{eqnarray}\label{sadd2}
I_{N}\left(\lambda_1,\lambda_2,\lambda_3\right)
& \sim &
\left(\frac{2\pi}{\frac{1}\kappa}\right)^\frac{N}{2}\;
\exp\left[-\frac{1}{\kappa}
\left(-S\left(t^{0};{\sf a},{\sf b},{\sf c}\right)\right)\right]
\nonumber
\\
&\times &
\left[{\rm Hess}\left(-S\left(t^0;{\sf a},{\sf b},{\sf c}\right)\right)\right]^{-\frac{1}{2}}.
\end{eqnarray}

On the other hand one can apply the Stirling formula to the r.h.s. of 
(\ref{Selberg}). 
This calculation has been performed by A.~Varchenko in \cite{Var}, who 
compared both asymptotics and 
got an interesting result, i.e., the following formulae:
\begin{eqnarray*}
\delta &=&\prod\limits_{k=0}^{N-1}
\frac{(k+1)^{k+1}{\sf c}^{k}({\sf a}+k{\sf c})^{k}
({\sf a}+k{\sf c})^{k}}
{({\sf a}+{\sf b}+(2N-k-2){\sf c})^{2N-k-2}},
\\
{\rm Hess}\left(-S\left(t^0;{\sf a},{\sf b},{\sf c}\right)\right) &=&
\prod\limits_{k=0}^{N-1}\frac{({\sf a}+{\sf b}+(2N-k-2){\sf c})^{3}}
{({\sf a}+k{\sf c})({\sf b}+k{\sf c})}.
\end{eqnarray*}

The integral (\ref{CN}), namely, 
\begin{eqnarray*}
C_{N}(b^2\rho,b^2\omega) &=& \int\limits_{0}^{1}{\rm d}u_{1}\ldots\int\limits_{0}^{1}{\rm d}u_{N}
{\rm e}^{-b^2 {\sf S}_{N}({\bf u},\rho,\omega)},
\\
{\sf S}_{N}({\bf u},\rho,\omega) &=&-\sum\limits_{i=1}^{N}\left(\rho\log u_i+\omega\log(u_i-1)\right)
-2\sum\limits_{1\leq i<j\leq N}
\log(u_j-u_i).
\end{eqnarray*}
is nothing but the Selberg integral. The saddle point approximation of $C_{N}(b^2\rho,b^2\omega)$ for
$b\to\infty$ yields the same as above, i.e.,\footnote{In fact, it is a bit easier to calculate, because here ${\sf c}=1$.}
$$
C_{N}(b^2\rho,b^2\omega)\;\sim\; \left(\frac{2\pi}{b^2}\right)^{\frac{N}{2}}
\frac{{\rm e}^{-b^2 {\sf S}_{N}(\rho,\omega)}}{\sqrt{{\rm Hess}({\sf S}_{N}(\rho,\omega))}},
$$
where ${\sf S}_{N}(\rho,\omega)\equiv{\sf S}_{N}\left({\bf u}^{\rm c},\rho,\omega\right)$ and ${\bf u}^{\rm c}$
is a solution of critical points eqs.: $\partial{\sf S}_{N}({\bf u},\rho,\omega)/\partial u_i =0$, $i=1,\ldots,N$.

As an example and for further uses 
(see appendix \ref{AppC}) let us calculate 
${\sf S}_{N}(\rho,\omega)$ in two simplest 
cases of $N=1$ and $N=2$. 
\begin{itemize}
\item\underline{$N=1$} Here,
$$
{\sf S}_{1}(u,\rho,\omega)=-\rho\log u -\omega\log(u-1)
=-\rho\log u -\omega\log(1-u)-i\pi\omega
$$
and
\begin{eqnarray}\label{C1}
C_{1}(b^2\rho, b^2\omega)=\int\limits_{0}^{1}{\rm d}u\,{\rm e}^{-
b^2 {\sf S}_{1}(u,\rho,\omega)}
&=&
{\rm e}^{b^2i\pi\omega}\int\limits_{0}^{1}{\rm d}u\, u^{b^2\rho}
\left(1-u\right)^{b^2\omega}\nonumber\\
&=&
{\rm e}^{b^2i\pi\omega}
\frac{\Gamma(b^2\rho+1)\Gamma(b^2\omega+1)}
{\Gamma(b^2\rho+b^2\omega+2)}.
\end{eqnarray}
The saddle point equation takes the form
$$
\frac{\rho}{u}+\frac{\omega}{u-1}\;=\;0
$$
and the solution is
$$
u^{\rm c}(\rho,\omega)=\frac{\rho}{\rho+\omega}.
$$
Therefore,
\begin{eqnarray}
{\sf S}_{1}\left(\rho,\omega\right)
\equiv{\sf S}_{1}\left(u^{\rm c}(\rho,\omega),\rho,\omega\right)
&=&-\rho\log\left(\frac{\rho }{\rho+\omega}\right)
-\omega\log\left(1-\frac{\rho}{\rho+\omega}\right)-i\pi\omega\nonumber
\\
&=&
-\rho\log\left(\frac{\rho }{\rho+\omega}\right)
-\omega\log\left(\frac{\omega}{\rho+\omega}\right)-i\pi\omega.
\end{eqnarray}
and
$$
C_{1}(b^2\rho, b^2\omega)\sim 
\left(\frac{2\pi}{b^2}\right)^{\frac{1}{2}}
\frac{{\rm e}^{-b^2 {\sf S}_{1}(\rho,\omega)}}
{\sqrt{\partial^{2}_{u}{\sf S}_{1}(\rho,\omega)}}.
$$
Applying Stirling's formula 
$\Gamma(z)=\sqrt{\frac{2\pi}{z}}\left(\frac{z}{\rm e}\right)^{z}
(1+O(1/z))$ to the r.h.s. of (\ref{C1}) one can get the same 
asymptotical 
behaviour of $C_{1}(b^2\rho, b^2\omega)$ for $b\to\infty$, namely, 
$$
\frac{\sqrt{2\pi}}{b}O(b^0)\exp(-b^2{\sf S}_{1}(\rho,\omega)).
$$

\item\underline{$N=2$} Here,
\begin{eqnarray*}
{\sf S}_{2}({\bf u},\rho,\omega) &=&-\rho\left(\log u_1 + \log
u_2\right)-\omega\left(\log(u_1-1)+
\log(u_2-1)\right)-2\log(u_2-u_1)
\end{eqnarray*}
and 
\begin{eqnarray*}
\frac{\partial {\sf S}_{2}}{\partial u_1}=0 &\Leftrightarrow&
\frac{\rho}{u_1}+\frac{\omega}{u_1-1}+\frac{2}{u_1-u_2}=0,\\
\frac{\partial {\sf S}_{2}}{\partial u_2}=0 &\Leftrightarrow&
\frac{\rho}{u_2}+\frac{\omega}{u_2-1}+\frac{2}{u_2-u_1}=0.
\end{eqnarray*}
The above equations have two pairs of analytical solutions. First,
\begin{eqnarray*}
u_{1}^{\rm c}(\rho,\omega) &=& \frac{\rho(\rho+\omega +2)-\sqrt{(\rho +1)(\omega +1)(\rho +\omega +1)}+\omega +1}
{(\rho +\omega +1)(\rho +\omega +2)},
\\
u_{2}^{\rm c}(\rho,\omega) &=& \frac{\rho(\rho +\omega +2)+\sqrt{(\rho +1) (\omega +1) (\rho +\omega +1)}+\omega +1}
{(\rho +\omega +1)(\rho +\omega +2)}
\end{eqnarray*}
and the second pair, which is a permutation of components of the first. So,
to calculate the critical value ${\sf S}_{2}\left({\bf u}^{\rm c},\rho,\omega\right)$ of 
${\sf S}_{2}\left({\bf u},\rho,\omega\right)$ we need to evaluate the latter on the critical solutions.
For instance,
$$
{\sf S}_{2}\left(\rho,\omega\right)\equiv{\sf S}_{2}\left(u_{1}^{\rm c}(\rho,\omega), u_{2}^{\rm c}(\rho,\omega),
\rho,\omega\right).
$$
\end{itemize}

\section{Numerical checks of the four-point relation}
\label{AppC}
In this appendix we test the conjectured relation (\ref{fW}) 
numerically 
in some special cases. Recall, that on the left side of the 
eq.~(\ref{fW})
occurs the $s$-channel classical four-point 
block on the sphere (cf.~eqs.~(\ref{fcl})-(\ref{fclS})), i.e.,
\begin{equation}
f(\delta_1,\ldots,\delta_4,\underline{\delta}\,|\,x)\;=\;
\left(\underline{\delta}-\delta_1-\delta_2\right)\log x+
\hat f(\delta_1,\ldots,\delta_4,\underline{\delta}\,|\,x).
\end{equation}
As already mentioned, the function 
$\hat f(\delta_1,\ldots,\delta_4,\underline{\delta}\,|\,x)$ is 
given by a power series in $x$, 
$$
\hat f(\delta_1,\ldots,\delta_4,\underline{\delta}\,|
\,x)=\sum\limits_{n=1}^{\infty}
x^n\hat f_{n}(\delta_1,\ldots,\delta_4,\underline{\delta}),
$$
where coefficients $\hat f_{n}(\delta_1,\ldots
\delta_4,\underline{\delta})$ are calculated 
directly from the asymptotic behaviour (\ref{cla1}) and the power 
expansion of the quantum block,
\begin{equation}\label{expansions}
\sum\limits_{n=1}^{\infty}
x^n\hat f_{n}(\delta_1,\ldots,\delta_4,\underline{\delta})
=\lim\limits_{b\to\infty}
\frac{1}{b^2}\log\left(1+\sum_{n=1}^\infty x^{n}
{\sf B}_{n}\!\left(\Delta_1,\ldots,
\Delta_4,\underline{\Delta},c\right)\right).
\end{equation}
Here, the central charge $c$ is given by 
(\ref{cVb}) and the relation (\ref{cl2})
between classical and quantum conformal weights is assumed.
It is also reasonable to assume that $0<|x|<1$, as it is dictated by 
the radial ordering of primary fields 
in the $s$-channel four-point function. 
For such $x$ both the quantum and classical four-point 
blocks are convergent. So, to get 
$\hat f_{n}(\delta_1,\ldots,\delta_4,\underline{\delta})$ 
one needs coefficients of the quantum block.
For lower orders of the expansion these coefficients can be easily 
computed directly from a 
definition. For instance (cf.~\cite{MMS}),
\begin{eqnarray}\label{B1}
{\sf B}_{1} &=& \frac{\left(\underline{\Delta}+\Delta_2-
\Delta_1\right) 
\left(\underline{\Delta}+\Delta_3-\Delta_4\right)}
{2\underline{\Delta}},
\\[5pt]
\label{B2}
{\sf B}_{2} &=&
\Big[\left(\underline{\Delta }-\Delta_1+\Delta_2\right)
\left(\underline{\Delta }-\Delta_1+\Delta_2+1\right) 
\left(\underline{\Delta }+\Delta_3-\Delta_4\right)\nonumber
\\
&&
\left(\underline{\Delta }+\Delta_3-\Delta_4+1\right)
\left(4 \underline{\Delta }+\frac{c}{2}\right)-6\underline{\Delta}
\left(\underline{\Delta }-\Delta_1+2 \Delta_2\right) \nonumber
\\
&&\left(\underline{\Delta }+\Delta_3-\Delta_4\right) 
\left(\underline{\Delta}+\Delta_3-\Delta_4+1\right)-6\underline{\Delta}
\left(\underline{\Delta }-\Delta_1+\Delta_2\right) \nonumber
\\
&&\left(\underline{\Delta }-\Delta_1+\Delta_2+1\right) 
\left(\underline{\Delta }+2 \Delta_3-\Delta_4\right)
+4 \underline{\Delta}
\left(\underline{\Delta }-\Delta_1+2 \Delta_2\right) \nonumber
\\
&&
\left(\underline{\Delta }+2 \Delta_3-\Delta_4\right) \left(2 \underline{\Delta }+1\right)\Big]
\Big[2 \underline{\Delta } \left(2 (c-5) \underline{\Delta}+16 \underline{\Delta }^2+c\right)\Big]^{-1}
\end{eqnarray}
and
\begin{eqnarray}
\label{B3}
{\sf B}_{3} &=& 
\left[2\underline{\Delta}(3\underline{\Delta}^2+c\underline{\Delta}-7\underline{\Delta}+2+c)\right]^{-1}\nonumber
\\
&\times & 
\Big[ 
(\underline{\Delta}+3\Delta_2-\Delta_1)
(\underline{\Delta}^2+3\underline{\Delta}+2)
(\underline{\Delta}+3\Delta_3-\Delta_4)\nonumber
\\
&-&2(\underline{\Delta}+3\Delta_2-\Delta_1)
(\underline{\Delta}+1)
(\underline{\Delta}+2\Delta_3-\Delta_4)
(\underline{\Delta}+\Delta_3-\Delta_4+2)\nonumber
\\
&+&(\underline{\Delta}+3\Delta_2-\Delta_1)
(\underline{\Delta}+\Delta_3-\Delta_4)
(\underline{\Delta}+\Delta_3-\Delta_4+1)
(\underline{\Delta}+\Delta_3-\Delta_4+2)\nonumber
\\
&-&2(\underline{\Delta}+2\Delta_2-\Delta_1)
(\underline{\Delta}+\Delta_2-\Delta_1+2)
(\underline{\Delta}+1)
(\underline{\Delta}+3\Delta_3-\Delta_4)\nonumber
\\
&+&(\underline{\Delta}+\Delta_2-\Delta_1)
(\underline{\Delta}+\Delta_2-\Delta_1+1)
(\underline{\Delta}+\Delta_2-\Delta_1+2)
(\underline{\Delta}+3\Delta_3-\Delta_4)\nonumber
\\
&+&2(\underline{\Delta}+2\Delta_2-\Delta_1)
(\underline{\Delta}+\Delta_2-\Delta_1+2)
(6\underline{\Delta}^3+9\underline{\Delta}^2
-9\underline{\Delta}+2c\underline{\Delta}^2+3c\underline{\Delta}+c)\nonumber
\\
&\times &
(\underline{\Delta}+2\Delta_3-\Delta_4)
(\underline{\Delta}+\Delta_3-\Delta_4+2)
\left[16\underline{\Delta}^2+2(c-5)\underline{\Delta}+c\right]^{-1}\nonumber
\\
&-&\left[16\underline{\Delta}^2+2(c-5)\underline{\Delta}+c\right]^{-1}
(\underline{\Delta}+2\Delta_2-\Delta_1)
(\underline{\Delta}+\Delta_2-\Delta_1+2)
(9\underline{\Delta}^2-7\underline{\Delta}+3c\underline{\Delta}+c)\nonumber
\\
&\times & 
(\underline{\Delta}+\Delta_3-\Delta_4)
(\underline{\Delta}+\Delta_3-\Delta_4+1)
(\underline{\Delta}+\Delta_3-\Delta_4+2)\nonumber
\\
&-&\left[16\underline{\Delta}^2+2(c-5)\underline{\Delta}+c\right]^{-1}
(\underline{\Delta}+\Delta_2-\Delta_1)
(\underline{\Delta}+\Delta_2-\Delta_1+1)
(\underline{\Delta}+\Delta_2-\Delta_1+2)\nonumber
\\
&\times &
(9\underline{\Delta}^2-7\underline{\Delta}+3c\underline{\Delta}+c)
(\underline{\Delta}+2\Delta_3-\Delta_4)
(\underline{\Delta}+\Delta_3-\Delta_4+2)\nonumber
\\
&+&
(\underline{\Delta}+\Delta_2-\Delta_1)
(\underline{\Delta}+\Delta_2-\Delta_1+1)
(\underline{\Delta}+\Delta_2-\Delta_1+2)\nonumber
\\
&\times &
(\underline{\Delta}+\Delta_3-\Delta_4)
(\underline{\Delta}+\Delta_3-\Delta_4+1)
(\underline{\Delta}+\Delta_3-\Delta_4+2)\nonumber
\\
&\times &
\frac{24\underline{\Delta}^2-26\underline{\Delta}+11c\underline{\Delta}+8c+c^2}
{12\left(16\underline{\Delta}^2+2(c-5)\underline{\Delta}+c\right)}
\Big].
\end{eqnarray}
For higher orders
the calculation of conformal block coefficients by inverting the Gram 
matrices become very 
laborious. A more efficient method based on recurrence relations for 
the coefficients can be used \cite{Z1,Z2,ZZ}.
Having in hands (\ref{B1})-(\ref{B3}) 
one can compute the classical block 
coefficients expanding the logarithm in (\ref{expansions}) into power 
series and then taking the limit of each term separately.
For $n=1,2,3$ one finds:
\begin{eqnarray*}
\hat f_{1} &=& \frac{\left(\underline{\delta}-\delta_1+
\delta_2\right) 
\left(\underline{\delta}+\delta_3-\delta_4\right)}
{2\underline{\delta}},
\\[5pt]
\hat f_{2} &=& \Big[13\underline{\delta}^5+
\left(-14\delta_1+18\delta_2+18\delta_3-14\delta_4-9\right)
\underline{\delta}^4
\\
&+&
(\delta_1^2-2\left(\delta_2+6\delta_3-10\delta_4-6\right)
\delta_1+\delta_2^2+\delta_3^2+
\delta_4^2
\\
&-&
12\delta_3+4\delta_2\left(5\delta_3-3\delta_4-3\right)
-2\delta_3\delta_4+12\delta_4)\underline{\delta}^3
\\
&-&3(\left(2\delta_3+2\delta_4+1\right)\delta_1^2
- 2(-\delta_3^2+2\left(\delta_4+1\right)\delta_3-
\delta_4\left(\delta_4+2\right)
\\
&+&\delta_2\left(2\delta_3+2\delta_4+1\right))\delta_1
+\left(\delta_3-\delta_4\right){}^2
\\
&+&
\delta_2^2 \left(2 \delta_3+2 \delta_4+1\right)
+2\delta_2(\delta_3^2-2\left(\delta_4-1\right)\delta_3
\\
&+&
\left(\delta_4-2\right)\delta_4))
\underline{\delta}^2+5\left(\delta_1
-\delta_2\right){}^2\left(\delta_3-\delta_4\right){}^2
\underline{\delta} 
+3\left(\delta_1-\delta_2\right){}^2
\left(\delta_3-\delta_4\right){}^2\Big]
\\
&\times &
\Big[16\underline{\delta}^3(4\underline{\delta}-3)\Big]^{-1}
\end{eqnarray*}
and
\begin{eqnarray*}
\hat f_{3} &=&
23\underline{\delta}^8+\left(-26\delta_1+38\delta_2+38\delta_3-26\delta_4-61\right)\underline{\delta}^7
+(3\delta_1^2+\left(-6\delta_2-21\delta_3+45\delta_4+76\right)\delta_1
\\
&+&
3\delta_2^2+3\delta_3^2+3\delta_4^2-100\delta_3+
\delta_2\left(45\delta_3-21\delta_4-100\right)-6\delta_3\delta_4+76\delta_4+30)\underline{\delta}^6
\\
&-&
3(\left(4 \delta_3+8\delta_4+5\right)\delta_1^2+\left(4\delta_3^2-12\delta_4\delta_3-23
\delta_3+8\delta_4^2+39\delta_4-2\delta_2\left(6\delta_3+6\delta_4+5\right)+16\right)
\delta_1
\\
&+&
5\delta_3^2+5\delta_4^2-16\delta_3-10\delta_3\delta_4+16\delta_4+\delta_2^2
\left(8\delta_3+4\delta_4+5\right)
\\
&+&\delta_2\left(8\delta_3^2-12\delta_4\delta_3+39\delta_3
+4\delta_4^2-23\delta_4-16\right))\underline{\delta}^5
\\
&+&(-5\left(\delta_3-\delta_4\right)\delta_1^3+\left(-5\delta_3^2+3\left(5\delta_2-10\delta_4+8\right)\delta_3+35
\delta_4^2+\left(48-15\delta_2\right)\delta_4+18\right)\delta_1^2
\\
&-&(5\delta_3^3-3\left(5\delta_4+8\right)\delta_3^2+3\left(5\delta_4^2+24\delta_4+18\right)\delta_3+15\delta_2^2
\left(\delta _3-\delta _4\right)
\\
&-&\delta_4\left(5\delta_4^2+48\delta_4+54\right)+6\delta_2(5\delta_3^2-2\left(5\delta_4-6\right) 
\delta_3+5\delta_4^2+12\delta_4+6))\delta_1
\\
&+&
18\left(\delta_3-\delta_4\right){}^2+5\delta_2^3\left(\delta_3-\delta_4\right)
+\delta_2^2\left(35\delta_3^2-6\left(5\delta_4-8\right)\delta_3-5\delta_4^2+24\delta_4
18\right)
\\
&+&\delta_2(5\delta_3^3+\left(48-15\delta_4\right)\delta_3^2
+3\left(5\delta_4^2-24\delta_4+18\right)\delta_3+
\delta_4\left(-5\delta_4^2+24\delta_4-54\right))) 
\underline{\delta}^4
\\
&+&
7(\delta_1-\delta_2)\left(\delta_3-\delta_4\right) 
(\left(2\delta_3+2\delta_4+1\right)\delta_1^2-(-2\delta_3^2+
\left(4\delta_4+3\right)\delta_3
\\
&-&\delta_4\left(2\delta_4+3\right)+\delta_2\left(4
\delta_3+4\delta_4+2\right)) 
\delta_1+\left(\delta_3-\delta_4\right){}^2+\delta_2^2
\left(2\delta_3+2\delta_4+1\right)
\\
&+&\delta_2\left(2\delta_3^2+\left(3-4\delta_4\right)\delta
_3+\delta_4\left(2\delta_4-3\right)\right))\underline{\delta}^3-3 
\left(\delta_1-\delta_2\right)\left(\delta _3-\delta _4\right) 
((3\delta_3^2-2\left(3\delta_4+2\right)\delta_3
\\
&+&
3\delta_4^2-4\delta_4-2)\delta_1^2+(-4\delta_3^2+\left(8\delta
_4+6\right) \delta _3-2 \delta _4 \left(2 \delta _4+3\right)+
\delta_2 
(-6\delta_3^2+4\left(3
\delta _4+2\right)\delta_3
\\
&-&
6\delta_4^2+8\delta_4+4))\delta_1-2\left(\delta
_3-\delta_4\right){}^2+\delta_2^2\left(3\delta_3^2-2\left(3\delta 
_4+2\right)\delta_3+3
\delta _4^2-4 \delta _4-2\right)
\\
&+&\delta_2\left(-4\delta_3^2+\left(8\delta_4-6\right)\delta_3+2
\left(3-2\delta_4\right)\delta_4\right))
\underline{\delta}^2-19\left(\delta_1-\delta
_2\right){}^3 
\\
&\times &
\left(\delta_3-\delta_4\right){}^3\underline{\delta}-6 
\left(\delta_1-\delta_2\right){}^3 \left(\delta_3
-\delta _4\right){}^3
\Big[48\underline{\delta}^5\left(4\underline{\delta}
^2-11\underline{\delta}+6\right)\Big]^{-1}.
\end{eqnarray*}

Turning to our task, let us stress again that
eq.~(\ref{fW}) was obtained as a result of considering the classical 
limit of the MMS relation (\ref{MMS}). Accordingly, the 
right side of the eq.~(\ref{fW}) is nothing but the saddle point 
approximation of the Dotsenko--Fateev integral that appears in 
the quantum identity (\ref{MMS}). The quantum MMS relation 
(\ref{MMS}) holds
provided that the relations (\ref{rel}) between 
parameters are assumed. In the eq.~(\ref{fW}) analogous 
relationships 
between the parameters are assumed, namely 
formulae (\ref{delta})-(\ref{d2}). 
Here, we check the relation (\ref{fW}) 
numerically in the simplest situation, i.e., for
\begin{equation}\label{paraS}
N_1=N_2=1
\;\;\;\;{\rm and}\;\;\;\;
\hat\alpha_1=\hat\alpha_2=\hat\alpha_3=b\eta.
\end{equation}
For this case, the Dotsenko--Fateev integral takes the form:
$$
{\rm e}^{b^2\left(2\eta^2\log x+2\eta^2\log (1-x)\right)}
\int\limits_{0}^{x}{\rm d}u_1
\int\limits_{0}^{1}{\rm d}u_2 {\rm e}^{-b^2 W(\eta,x,u_1,u_2)},
$$
where
\begin{eqnarray*}
W(\eta,x,u_1, u_2) &=&
-2 \log (u_2-u_1)
-2\eta 
\Big(\log u_1+\log u_2
\\
&+&\log(u_1-x)+\log(u_2-x)
+\log(u_1-1)+\log(u_2-1)\Big)\\
&=&
-2 \log (u_2-u_1)-2\eta\Big(\log u_1+\log u_2
\\
&+&\log(x-u_1)+\log(u_2-x)
+\log(1-u_1)+\log(1-u_2)+3i\pi\Big).
\end{eqnarray*}
The saddle point equations read as follows:
\begin{eqnarray}\label{sadd1}
\frac{\partial W}{\partial u_1}=0 &\Leftrightarrow&
2\eta\left(\frac{1}{u_1-1}+\frac{1}{u_1}+\frac{1}{u_1-x}\right)+\frac{2}{u_1-u_2}=0,\\
\label{sadd2}
\frac{\partial W}{\partial u_2}=0 &\Leftrightarrow&
2\eta\left(\frac{1}{u_2-1}+\frac{1}{u_2}+\frac{1}{u_2-x}\right)+\frac{2}{u_2-u_1}=0.
\end{eqnarray}
For given $\eta$ and $x$, using basic tools available in Mathematica, 
such as the {\sf NSolve} procedure, 
one can find {\it real} numerical solutions to the above 
equations.\footnote{Here we consider {\it only} real solutions.}
A list of these solutions is provided below.
Hence, for $b\to\infty$
$$
\int\limits_{0}^{x}{\rm d}u_1
\int\limits_{0}^{1}{\rm d}u_2 {\rm e}^{-b^2 W(\eta,x,u_1,u_2)}
\;\sim\;
\frac{2\pi}{b^2}
\frac{{\rm e}^{-b^2 W(\eta,x,u_{1}^{\rm c},u_{2}^{\rm c})}}
{\sqrt{{\rm Hess}\left(W(\eta,x,u_{1}^{\rm c},u_{2}^{\rm c})\right)}}.
$$
Thus, in this case, the asymptotical behaviour of the left-hand side of the four-point MMS relation (\ref{MMS}) is
\begin{equation}\label{lhs}
\frac{2\pi}{b^2}\left[{\rm Hess}\left(W(\eta,x,u_{1}^{\rm c},u_{2}^{\rm c})\right)\right]^{-\frac{1}{2}}
{\rm e}^{-b^2 \left(W(\eta,x,u_{1}^{\rm c},u_{2}^{\rm c})-2\eta^2\log x-2\eta^2\log(1-x)\right)}.
\end{equation}
For (\ref{paraS}), the r.h.s. of (\ref{MMS}) 
in the limit $b\to\infty$ yields
\begin{eqnarray}\label{rhs}
&&C_{1}\left(b^2 2\eta, b^2 2\eta\right)\,
C_{1}\left(b^2 2(2\eta+1), b^2 2\eta\right)\,
x^{\underline{\Delta}-2\tilde\Delta}\,
{\sf B}\!\left(\tilde\Delta,\tilde\Delta,\tilde\Delta,
\Delta_4,\underline{\Delta},c\,|\,x\right) \;\sim\;\nonumber
\\[5pt]
&&\hspace{-30pt}
\left(\frac{2\pi}{b^2}\right)^{\frac{1}{2}}
\frac{{\rm e}^{-b^2 {\sf S}_{1}(2\eta,2\eta)}}
{\sqrt{\partial^{2}_{u}{\sf S}_{1}(2\eta,2\eta)}}\,
\left(\frac{2\pi}{b^2}\right)^{\frac{1}{2}}
\frac{{\rm e}^{-b^2 {\sf S}_{1}(2(2\eta+1),2\eta)}}
{\sqrt{\partial^{2}_{u}{\sf S}_{1}(2(2\eta+1),2\eta)}}
\,{\rm e}^{b^2\left[\left(\underline{\delta}-2\tilde\delta\right)
\log x+{\hat f}(\tilde\delta,\tilde\delta,\tilde\delta,
\delta_4,\underline{\delta}|x)\right]},
\end{eqnarray}
where the quantum block parameters are given by 
\begin{eqnarray*}
\tilde\Delta &=& b\eta\left(b\eta+\tfrac{1}{b}-b\right),
\quad\quad
\Delta_4\;=\;\left(2b+3b\eta\right)\left(2b+3b\eta+\tfrac{1}{b}-
b\right),\\
\underline\Delta &=& \left(b+2b\eta\right)\left(b+2b\eta+\tfrac{1}
{b}-b\right),
\quad\quad
c\;=\;1-6\left(b-\tfrac{1}{b}\right)^2
\end{eqnarray*}
while the classical conformal weights are parameterized as 
follows
\begin{equation}\label{del}
\tilde\delta\;=\;\eta\left(\eta-1\right),
\quad\quad
\delta_{4}\;=\;\left(2+3\eta\right)\left(1+3\eta\right),
\quad\quad
\underline\delta\;=\;2\eta\left(2\eta+1\right).
\end{equation}
Finally, comparing both asymptotics, i.e., 
expressions (\ref{lhs}) and (\ref{rhs}), and taking the limit
$b\to\infty$ 
one gets\footnote{In the limit $b\to\infty$ vanish additional 
terms of the 
form $\frac{1}{b^2}\log(\frac{2\pi}{b^2}{\cal O}(b^0))$.}
\begin{eqnarray}\label{Class4ptRel}
&&W(\eta,x,u_{1}^{\rm c},u_{2}^{\rm c})-2\eta^2\log 
x-2\eta^2\log(1-x)\nonumber
\;=\;\\
&&\hspace{30pt}
{\sf S}_{1}(2\eta,2\eta)+{\sf S}_{1}(2(2\eta+1),2\eta)
-(\underline{\delta}-2\tilde\delta)\log x 
-{\hat f}(\tilde\delta,\tilde\delta,\tilde\delta,
\delta_4,\underline{\delta}\,|\,x).
\end{eqnarray}

\subsubsection*{Numerics}
The relation (\ref{Class4ptRel}) we check numerically for
$\eta=10,100$ and  $x=0.1,0.25,0.5,0.75$.
In these calculations the selected values of the parameter $\eta$ 
are limited only by 
the requirement that the corresponding values of the classical 
conformal weights (\ref{del}) are positive.
Numerical calculations can be made for a much larger set of 
parameters, which we did. However, this leads to the same 
conclusions as spelled out below.

First, we calculate the left-hand side of (\ref{Class4ptRel}) for 
the above given values of parameters $\eta$ and $x$, and all 
corresponding 
real critical points $\lbrace u_{1}^{\rm c},u_{2}^{\rm c}\rbrace$ 
obeying (\ref{sadd1})-(\ref{sadd2}).

\medskip\noindent
$\bullet\;\underline{\eta=10}$
\begin{itemize}
\item[]
\centering
\begin{tabular}{ |l|l| }
  \hline
  \multicolumn{2}{|c|}{$x=0.1$} \\
  \hline
  \hspace{1.2cm}$\lbrace u_{1}^{\rm c},u_{2}^{\rm c}\rbrace$ & {\rm l.h.s.}\;{\rm of}\;(\ref{Class4ptRel})\\
  \hline
  $\{0.056576, 0.040983\}$ & $732.582-257.611i$\\
  $\{0.040983, 0.056576\}$ & $732.582-251.327i$ \\
  $\{0.694717, 0.048494\}$ & $644.713-194.779i$ \\
  $\{0.048494, 0.694717\}$ & $644.713-188.496i$ \\
  $\{0.622355, 0.736874\}$ & $569.705-125.664i$ \\
  $\{0.736874, 0.622355\}$ & $569.705-131.947i$ \\
  \hline 
\end{tabular}
\item[]
\centering 
\begin{tabular}{ |l|l| }
  \hline
  \multicolumn{2}{|c|}{$x=0.25$} \\
  \hline
   \hspace{1.2cm}$\lbrace u_{1}^{\rm c},u_{2}^{\rm c}\rbrace$ & {\rm l.h.s.}\;{\rm of}\;(\ref{Class4ptRel})\\
  \hline
  $\{0.136106, 0.097503\}$ & $513.788-257.611i$\\
  $\{0.097503, 0.136106\}$ & $513.788-251.327i$ \\
  $\{0.114959, 0.725611\}$ & $468.673-188.496i$ \\
  $\{0.763901, 0.661919\}$ & $434.612-131.947i$ \\
  $\{0.661919, 0.763901\}$ & $434.612-125.664i$ \\
   \hline
\end{tabular}
\item[]
\centering
\begin{tabular}{ |l|l| }
  \hline
  \multicolumn{2}{|c|}{$x=0.5$} \\
  \hline
  \hspace{1.2cm}$\lbrace u_{1}^{\rm c},u_{2}^{\rm c}\rbrace$ & {\rm l.h.s.}\;{\rm of}\;(\ref{Class4ptRel})\\
  \hline
  $\{0.823332, 0.749419\}$  & $404.829-131.947i$\\
  $\{0.749419, 0.823332\}$  & $404.829-125.664i$ \\
  $\{0.176668, 0.250580\}$  & $404.829-251.327i$ \\
  $\{0.250580, 0.176668\}$  & $404.829-257.611i$ \\
  $\{0.793369, 0.206631\}$  & $399.71-194.779i$ \\
  $\{0.206631, 0.793369\}$  & $399.71-188.496i$ \\
   \hline
\end{tabular}
\item[]
\centering 
\begin{tabular}{ |l|l| }
  \hline
  \multicolumn{2}{|c|}{$x=0.75$} \\
  \hline
  \hspace{1.2cm}$\lbrace u_{1}^{\rm c},u_{2}^{\rm c}\rbrace$ & {\rm l.h.s.}\;{\rm of}\;(\ref{Class4ptRel})\\
  \hline
  $\{0.863894, 0.902497\}$ & $513.788-125.664i$\\
  $\{0.902497, 0.863894\}$ & $513.788-131.947i$ \\
  $\{0.274389, 0.885041\}$ & $468.673-188.496i$ \\
  $\{0.885041, 0.274389\}$ & $468.673-194.779i$ \\
  $\{0.338081, 0.236098\}$ & $434.612-257.611i$ \\
  $\{0.236098, 0.338081\}$ & $434.612-251.327i$ \\
   \hline
\end{tabular}
\end{itemize}

$\bullet\;\underline{\eta=100}$
\begin{itemize}
\item[]
\centering
\begin{tabular}{ |l|l| }
  \hline
  \multicolumn{2}{|c|}{$x=0.1$} \\
  \hline
  \hspace{1.2cm}$\lbrace u_{1}^{\rm c},u_{2}^{\rm c}\rbrace$ & {\rm l.h.s.}\;{\rm of}\;(\ref{Class4ptRel})\\
  \hline
  $\{0.051189, 0.046204\}$ & $50587.3 -2519.56i$\\
  $\{0.046204, 0.051189\}$ & $50587.3 -2513.27i$ \\
  $\{0.685683, 0.048667\}$ & $49782.2 -1891.24i$ \\
  $\{0.048667, 0.685683\}$ & $49782.2 -1884.96i$ \\
  $\{0.665947, 0.702308\}$ & $48994.4 -1256.64i$ \\
  $\{0.702308, 0.665947\}$ & $48994.4 -1262.92i$ \\
  \hline 
\end{tabular}
\item[]
\centering 
\begin{tabular}{ |l|l| }
  \hline
  \multicolumn{2}{|c|}{$x=0.25$} \\
  \hline
   \hspace{1.2cm}$\lbrace u_{1}^{\rm c},u_{2}^{\rm c}\rbrace$ & {\rm l.h.s.}\;{\rm of}\;(\ref{Class4ptRel})\\
  \hline
  $\{0.110100, 0.122433\}$ & $35204.3-2513.27i$ \\
  $\{0.122433, 0.110100\}$ & $35204.3-2519.56i$ \\
  $\{0.116077, 0.718001\}$ & $34809.3-1884.96i$ \\
  $\{0.718001, 0.116077\}$ & $34809.3-1891.24i$ \\
  $\{0.732896, 0.700490\}$ & $34430. -1262.92i$ \\
  $\{0.700490, 0.732896\}$ & $34430. -1256.64i$ \\
   \hline
\end{tabular}
\item[]
\centering
\begin{tabular}{ |l|l| }
  \hline
  \multicolumn{2}{|c|}{$x=0.5$} \\
  \hline
  \hspace{1.2cm}$\lbrace u_{1}^{\rm c},u_{2}^{\rm c}\rbrace$ & {\rm l.h.s.}\;{\rm of}\;(\ref{Class4ptRel})\\
  \hline
  $\{0.77666,  0.800211\}$ & $28948.1-1256.64i$\\
  $\{0.19979,  0.22334\}$  & $28948.1-2513.27i$ \\
  $\{0.22334,  0.199789\}$ & $28948.1-2519.56i$ \\
  $\{0.210845, 0.789155\}$ & $28940.7-1884.96i$ \\
  \hline
\end{tabular}
\item[]
\centering 
\begin{tabular}{ |l|l| }
  \hline
  \multicolumn{2}{|c|}{$x=0.75$} \\
  \hline
  \hspace{1.2cm}$\lbrace u_{1}^{\rm c},u_{2}^{\rm c}\rbrace$ & {\rm l.h.s.}\;{\rm of}\;(\ref{Class4ptRel})\\
  \hline
  $\{0.889899, 0.877567\}$ & $35204.3-1262.92i$\\
  $\{0.883922, 0.281999\}$ & $35204.3-1256.64i$ \\
  $\{0.883922, 0.281999\}$ & $34809.3-1891.24i$ \\
  $\{0.281999, 0.883922\}$ & $34809.3 -1884.96i$ \\
  $\{0.267103, 0.29951\}$  & $34430.-2513.27i$ \\
  $\{0.29951,  0.267103\}$ & $34430.-2519.56i$ \\
   \hline
\end{tabular}
\end{itemize}
Looking at the tables above we see that appear here real solutions
to the equations (\ref{sadd1})-(\ref{sadd2}) such that permutations 
(transpositions) of which are also solutions.
This is a known property 
of solutions of such type of (Bethe's) equations.

For the same values of $\eta$ and $x$ 
as above we calculate the 
numerical values of the r.h.s. of the relation (\ref{Class4ptRel}). 
The results are 
collected in the Table \ref{table:3}, 
where in the third and fourth columns are 
the values of the r.h.s. of the relation 
(\ref{Class4ptRel}) approximated by the 
second and third partial sums of the classical block,  respectively.
\begin{table}[h!]
\centering
\begin{tabular}{||c c c c||} 
\hline
$\eta$ & $x$ & {\rm r.h.s.}\;{\rm of}\;(\ref{Class4ptRel})
\,($\hat f\approx$ 2nd part.~sum) & {\rm r.h.s.}\;{\rm of}\;
(\ref{Class4ptRel})\,($\hat f\approx$ 3rd part.~sum)
\\ [0.5ex] 
 \hline\hline
 10 & $0.1$  & $644.627-125.664i$ & $644.706-125.664i$ \\
 \hline
 10 & $0.25$ & $467.141-125.664i$ & $468.38-125.664i$ \\ 
 \hline
 10 & $0.5$  & $383.442-125.664i$ & $393.358-125.664i$ \\
 \hline
 10 & $0.75$ & $383.724-125.664i$ & $417.192-125.664i$ \\
 \hline
 100 & $0.1$ & $49774.8-1256.64i$ & $49781.6-1256.64i$ \\
 \hline
 100 & $0.25$ & $34678.1-1256.64i$ & $34784.3-1256.64i$ \\
 \hline
 100 & $0.5$  & $27550.6-1256.64i$ & $28400.5-1256.64i$ \\ 
 \hline
 100 & $0.75$ & $27566.7-1256.64i$ & $304351.1-1256.64i$ \\ [1ex] 
 \hline
\end{tabular}
\caption{}
\label{table:3}
\end{table}

Ultimately, comparing the data in all the tables  
it can be seen that 
there is one real critical point, such that 
$0<u_{1}^{\rm c}<x<u_{2}^{\rm c}<1$,
for which the real parts of both sides of the relation
(\ref{Class4ptRel}) are almost the 
same for $x$ near $0$. These values start to differ from each other 
when we move away from $x=0$, which is due to the ``poorer'' 
convergence of the classical block. Unfortunately, the imaginary 
parts of both sides of the relation 
(\ref{Class4ptRel}) differ a bit.

\acknowledgments
The research of MRP has been supported by the 
Polish National Science Centre under grant no. ${\rm 2020/37/B/ST3/01471}$.


%
%
%
%
%
%
%
\end{document}